\documentclass[preprintnumbers,prd,floatfix,superscriptaddress,nofootinbib,notitlepage]{revtex4-1}
\pdfoutput=1
\usepackage{graphicx,epsfig}
\usepackage{amsmath}
\usepackage {amssymb}
\usepackage {longtable}
\usepackage{multirow}
\usepackage{dcolumn}
\usepackage{bm}
\usepackage{amsfonts}
\usepackage{subfigure}
\usepackage{color}
\usepackage{relsize}

\usepackage{enumitem}
\begin{document}

\title{Penrose and super-Penrose energy extraction from a
Reissner-Nordstr\"om black hole spacetime with a cosmological constant
through the BSW mechanism: Full story}

\author{Duarte Feiteira}
\email{duartefeiteira@tecnico.ulisboa.pt}
\affiliation{Centro de Astrof\'{\i}sica e Gravita\c c\~ao  - CENTRA,
Departamento de F\'{\i}sica, Instituto Superior T\'ecnico - IST,
Universidade de Lisboa - UL, Avenida Rovisco Pais 1, 1049-001
Lisboa, Portugal}

\author{Jos\'e P. S. Lemos}
\email{joselemos@ist.utl.pt}
\affiliation{Centro de Astrof\'{\i}sica e Gravita\c c\~ao  - CENTRA,
Departamento de F\'{\i}sica, Instituto Superior T\'ecnico - IST,
Universidade de Lisboa - UL, Avenida Rovisco Pais 1, 1049-001
Lisboa, Portugal}

\author{Oleg B. Zaslavskii}
\email{zaslav@ukr.net}
\affiliation{Department of Physics and Technology,
Kharkov V.~N.~Karazin National
University, 4 Svoboda Square, 61022 Kharkov, Ukraine}

\begin{abstract}

The Penrose process, a process that transfers energy from a black
hole to infinity, together with the BSW mechanism, a mechanism that
uses collisions of ingoing particles at the event horizon of a black
hole to locally produce large amounts of energy, is studied in a
combined description for a $d$ dimensional extremal
Reissner-Nordstr\"om black hole spacetime with negative, zero, or
positive cosmological constant, i.e., for an asymptotically anti-de
Sitter (AdS), flat, or de Sitter (dS) spacetime.
This blending of the Penrose process with BSW collisions is an
instance of a collisional Penrose process.
In an electrically charged extremal Reissner-Nordstr\"om black hole
background, in the vicinity of the event horizon, several types of
radial collisions between electrically charged particles can be
considered, the most interesting one is between a critical particle,
i.e., a particle that has its electric charge adjusted in a specific
way to the other relevant parameters, and a usual particle, as it
gives a divergent center of mass frame energy locally. A divergent
center of mass frame energy at the point of collision is a favorable
condition to extract energy from the black hole, but not sufficient,
since, e.g., the product particles might go down the hole. So, to
understand whether energy can be extracted or not in a Penrose
process, we investigate in detail a collision between ingoing
particles 1 and 2, from which particles 3 and 4 emerge, with the
possibility that particle 3 can carry the energy extracted far out
from the black hole horizon, i.e., there is a high Killing energy
transported by particle 3. One finds that the mass, the energy, the
electric charge, and the initial direction of motion of particle 3 can
have different values, depending on the collision internal process
itself. But, the different possible values of the the parameters of
the emitted particle 3 lie within some range, and moreover the energy
of particle 3 can, in some cases, be arbitrarily high but not
infinite, characterizing a super-Penrose process.
It is also shown that particle 4 has negative energy, as required
in a Penrose process, living in its own electric ergosphere while it
exists, i.e., before being engulfed by the event horizon.
For zero cosmological constant we find that the results do not depend
on the number of dimensions, but they do for negative and positive
cosmological constant. The value of the cosmological constant also
introduces differences in the lower bound for the energy extracted.

\end{abstract}
\maketitle

\section{Introduction}
\label{sec:intro}

Processes that can extract energy from a black hole are relevant on
several counts. They have astrophysical import whereby matter or waves
passing in the vicinity of an astrophysical black hole succeed in
getting energy out of it. They have physical relevance since black
holes can exist in all scales, in particular in microscales, and thus
can be of use, in principle,
 as an additional power source
in an advanced technology scheme.  Whereas
astrophysical black holes arise through gravitational collapse of
large quantities of matter, micro black holes with very small radii,
of the order of the neutron radius or even lower, can appear, for
instance, through pair creation in strong field regimes, or by
smashing matter fragments of exceedingly high energy against each
other.  Thus, the detail knowledge of all the possible mechanisms that
can be employed to extract energy from a black hole of any size
worth of pursuit.

Extraction of energy from a black hole started by the observation that
the ergoregion of a rotating black hole contains states with negative
energy with respect to infinity, and thus it is possible to somehow
deposit energy far away at the expense of the black hole energy.
This is the Penrose process \cite{penrose_book,penrosefloyd}.  In the
original process, it was proposed that a particle decays into two new
particles inside the ergoregion, and one of the created particles
carries an extra energy to infinity, with the energy source being the
black hole angular momentum, this process being thus a decayment
Penrose process.  A static electrically charged black hole
can allow as well for a Penrose process, the energy extracted coming
now from the electric ergosphere of the black hole \cite{denardo}.
A general study of the electric Penrose process
where electrically charged particles suffer a decayment in
an electrically charged black hole background
has been done in \cite{zaslavskii2024}.

Another process that yields a possible extraction of energy from a
black hole is the BSW mechanism \cite{bsw}.  The
original BSW mechanism
involves the collision at the horizon of an extremal rotating Kerr
black hole of two ingoing test particles, one of the particles being
critical, i.e., its angular momentum being critically tuned for the
effect to occur, so that generation of unbounded center of mass
energies would arise. Thus, we define BSW mechanism as a mechanism
involving two particles moving inwards and colliding at the horizon of
a black hole, or in the vicinity of it, where an unlimited center of
mass energy is produced locally.
The derivation of this mechanism motivated further study
to understand which type of black holes can provide the
effect and
whether the emission of highly energetic or supermassive
particles is possible after such a collision, and where the colliding
particles can be of any type, including dark matter and electrically
charged particles.
In this way,
it was shown that instead of an extremal black hole, a
nonextremal black hole could be used if the parameters for the particle
were properly adjusted at the point of the collision \cite{grib}.
It was also proved in \cite{zaslavskii_1007} that the BSW effect is
generic, i.e., it is due to the general properties of a black hole
horizon, and thus, many results that followed
can be seen as particular cases of this
feature.

As is typical in black hole physics, if some phenomenon exists for
rotating black holes and neutral particles, it is quite natural that
it also exists for electrically charged black holes and electrically
charged particles, and consequently it was soon proposed that an extremal
electrically charged Reissner-Nordstr\"om black hole, could also
yield, by the collision of two
ingoing electrically charged particles, with
one of them critical, a divergent center of mass frame energy,
in what is the electrically charged version of the BSW effect
\cite{zaslavskii_2010}.
Several other important
advancements on the BSW effect appeared as we now
mention.
By using the innermost stable circular orbit up to the extremal Kerr
state it was confirmed in \cite{haradakimura1} that high center of
mass energies can be created at the horizon and highly energetic
particles are subsequently emitted at the point of collision.
A geometric and general explanation, based on spacetime properties of
systems composed of rotating black holes and matter systems, pictured
the effect as attributes of null and timelike vectors in the vicinity
of the future event horizon \cite{zaslavskii1011}.
The creation of unbounded center of mass energies at the point of
collision was exhibit to persist with neutral particles in Kerr black
hole spacetimes in a cosmological constant setting \cite{li}.
The collision of two generic geodesic particles around a Kerr black
hole was further explored to produce unlimited center of mass energies
at the horizon in \cite{haradakimua1102}.

Clarifications and interesting applications of the BSW effect
kept emerging in the literature.
For instance,
a kinematic explanation of the effect was developed in
\cite{zaslavskii1104} in that a collision between a particle with
velocity tending to the velocity of light and a particle with a
velocity smaller than the velocity of light, both velocities seen in a
locally nonrotating frame at the horizon, produce an unlimited center
of mass energy.
An anti-de Sitter (AdS) background for extremal electrically charged
rotating cylindrical black holes was the scene to inspect the existence
of the BSW effect \cite{saidadami1105}.
For dirty rotating black holes, i.e., black holes with surrounding
matter, it was found that in the collisions of particles near the
horizon, the energy of the particles scales with a power of the
inverse surface gravity of the black hole \cite{zaslavskii2012}.
That ultrahigh energies at the center of mass scale
with some power of
the inverse surface gravity of the black hole was also uncovered by
considering, near the horizon of a Reissner-Nordstr\"om black hole
with negative cosmological constant, that one of the electrically
charged colliding particles is critical and at rest, which is possible
due to the electric repulsion on one hand, and the gravitational and
the extra attraction from the negative cosmological constant on the other hand
\cite{zaslavskii_lambda}.
Additionally, it was realized in \cite{zaslavskii2} that
electrically charged particles in radial motion that collide at the
horizon of an extremal Reissner-Nordstr\"om black hole can have debris
with no upper Killing energy bounds in
contrast to the finite energy of the ejected particles from a rotating
black hole.

The BSW effect has also been studied in higher and lower dimensions,
and has been implemented in several possible scenarios. 
In higher-dimensional spacetimes, the effect has been tested
specifically in extremal Myers-Perry black holes
\cite{kimura2}.
The analysis of collisions for higher-dimensional rotating black holes
was revisited in \cite{zaslavskii1409} establishing that high center
of mass energies can be produced.
A review of the BSW effect with emphasis on the collision of particles
in an extremal Kerr black hole was produced in \cite{haradakimura}.
The joint effect of rotation and electric charge was taken into
account for the extremal Kerr-Newman black hole, and in particular, a
noticeable BSW effect was found when simultaneously the angular
momentum of the particle is very large and the black hole charge is
very small \cite{hejda1}.
Lower-dimensional spacetimes were the target of a work for
the implementation of the effect, namely,
the BTZ rotating extremal black hole which is
a solution of 3-dimensional general relativity with negative
cosmological constant \cite{tsukamoto1705}.
An additional 3-dimensional rotating black hole, solution of a
topologically massive gravity, acts, as expected, as a particle
accelerator as discussed in \cite{becar}.
Several types of classification for different BSW 
scenarios have been performed in
\cite{ovcharenkozaslavskii2304,ovcharenkozaslavskii2402,
ovcharenkozaslavskii2405}.

One can combine the Penrose process and the BSW effect to
get a collisional Penrose process.
In the Penrose process one uses the negative states
existing in the ergosphere to extract energy of the black
hole which is then transferred to infinity.
The first instance found
for a physical Penrose process was through
particle decay in the ergosphere, but one can think
of other processes, a distinct one is indeed through
particle collisions, such as a BSW collision.
For collisions, the energy extracted
in a Penrose process from a rotating black hole can be
somewhat enhanced.
A relevant development and clarification of the BSW effect
demonstrated that although the center of mass energy grows
without bound
in a collision near or at an extremal Kerr horizon, the energy of the
particles at infinity, i.e., the Killing energy, is finite
and not large
\cite{bejger,zaslavskii1,harada}.
Still in the rotating black hole case, the relation
between the energy in the center of mass frame of BSW colliding particles
and the net Penrose energy extracted was uncovered in
\cite{zaslavskii3}.
Improvements and refinements of the collisional Penrose process
have appeared. For instance,
particle collisions in the axis of symmetry of a Kerr-Newman black
hole along with the possibility of energy extraction were described in
\cite{hejda2}.
For electrically charged colliding particles it was established that
even in a nonextremal Reissner-Nordstr\"om black hole background, the
emergent particles can carry arbitrarily large Killing energy to
infinity, thus yielding a super-Penrose process of real energy
extraction from a black hole \cite{zaslavskii6}, and confirming
previous results for
extremal electrically charged black hole backgrounds.
This extraction of arbitrary large amounts of energy
in particle collisions in electric charged black holes
contrasts with
the finite and moderate energy extraction from
the ejected particles of a rotating
black hole.
In addition,
a general and encompassing treatment of the BSW effect in the
equatorial plane of Kerr-Newman black holes where unbounded energies
taken by the particle debris to infinity can happen, yielding thus
real energy extraction from the black hole, was reported in
\cite{new,newnew}.
One could continue to enumerate processes that extract energy
from a black hole.
In a collision of two particles, one ingoing
the other outgoing, and so different from a BSW
collision in which the two colliding particles
are ingoing, it was shown that such a collisional Penrose
process could have an energetic extraction enhancement
by a factor close to 14 \cite{schnittman1410}, see also
\cite{ovcharenkozaslavskii2404}.
In a extended version of the decayment Penrose process,
one employs a mirror at some
radius surrounding the black hole,
to obtain multi processes and
so more energy extraction, leading possibly
to a black hole bomb, see \cite{flz}
for the electrically charged case.
Other processes,
being somewhat different
in character from the BSW effect or the Penrose process, can still
extract energy from a collision of particles.
These processes include particle
collisions in black hole spacetimes with additional physical features,
in spacetimes without black holes, and in other special spacetimes.
One can also extract energy from black holes using waves
in the superradiance phenomenon.

It is our aim to study the Penrose process in conjunction with the BSW
effect, i.e., the collision Penrose process, of black hole energy
extraction due to the collision of two electrically charged particles
at the event horizon of an extremal Reissner-Nordstr\"om black hole in
a background with a cosmological constant in generic $d$ dimensions in
a unified way. The cosmological constant will be allowed to have
negative values, in which case the spacetime is asymptotically AdS, to
have zero value, in which case the spacetime is asymptotically flat,
and to have positive values, in which case the spacetime is
asymptotically de Sitter (dS), in this latter case there is also the
cosmological horizon, but it does not play any major role in our
analysis. The number of dimensions obeys $d\geq 4$, and thus some of
the results already obtained for the particular case $d=4$ are
recovered.  One can list some of the reasons why these spacetimes are
relevant to study.
Reissner-Nordstr\"om black holes in isolation can in some instances
keep their own electric charge, although when surrounded by some
medium they are exposed to be completely discharged.
Asymptotically AdS spacetimes possess many symmetries and are the base
of some extended theories of gravity such as supergavity and string
theory.  Asymptotically flat spacetimes are the appropriate
environment in the study of a sufficiently large neighborhood of any
cosmic environment, such as the environment surrounding a black hole,
since for it the asymptotic structure is approximately flat.
Asymptotically dS spacetimes have implications in fundamental theories
and can be used describe the universe at large.  The interest in
studying spacetimes with $d$ generic dimensions comes from the
possibility of understanding what is peculiar to $d=4$ and what is
generic, and from the fact that several possible suitable theories
live consistently in spacetimes with higher dimensions, which makes
physical effects in these dimensions worth pursuing.

The work is organized as follows.
In Sec.~\ref{sec:equations}, the $d$-dimensional Reissner-Nordstr\"om
black hole spacetime in a cosmological constant background,
nonextremal and extremal, is introduced along with the important
horizon radii. The equations describing electrically charged
particle motion in such a spacetime are presented
and the electric ergosphere is defined.
In Sec.~\ref{sec:ECM}, the definitions and necessary conditions for
energy extraction from particle collisions are given, in particular,
we present the definition of critical, near-critical, and usual
particles, we make an analysis of the energy of the
BSW collisions at the
center of mass frame,  give the prerequisites for the particles to
reach the horizon, and show the only type of collision
that yields an unbounded center of mass energy is the collision
between a critical and a usual particle.
In Sec.~\ref{sec:E_emitted}, we examine in detail the energy extraction
from a collisional Penrose process.
For that, we uncover
the BSW collision between a critical particle that goes into the
black hole and a usual particle that also goes in, from which emerges
a near critical particle that goes out and a usual particle
with negative energy that goes
in, with the energy of the emitted particle, i.e., the possible energy
extracted in the Penrose process,
being established in terms of lower and upper bounds.
It is shown that super-Penrose processes can occur.
We add a discussion on the dependence of the process on the
cosmological constant and on the spacetime dimension $d$ and
make further comments.
In Sec.~\ref{sec:concl}, we conclude.
In the Appendix \ref{sec:app_app} we derive the center of mass
energy expression from the equations obtained for the energy of the
emitted particle, and uncover some details of the
energetics of the particle with negative energy
that falls in after the collision.

\vfill

\section{Line element, black hole horizon radii,
equations of motion for the particles, and the electric ergosphere}
\label{sec:equations}

\subsection{Line element}

In this work, it is provided a general analysis of the
collisional Penrose process, i.e.,
the combination of BSW effect with the Penrose
process of extraction of energy, in a
Reissner-Nordstr\"om black hole
background, with a negative, zero,
and positive cosmological constant in $d$ dimensions
with $d\geq4$.
Thus, we consider
the Reissner-Nordstr\"om-Tangherlini,
or simply Reissner-Nordstr\"om,
line
element 
for the interval $s$.
Generically, the spacetime
line element is written as
$ds^2=g_{ab}dx^adx^b$, where $g_{ab}$ is the metric
and the $dx^a$ are the coordinate
components of the infinitesimal  displacements,
with $a,b$ running over the time and spatial coordinates.
In usual spherical coordinates
$x^a=(t,r,\theta_1,\cdots,\theta_{d-2})$,
one can write 
$ds^2=g_{tt}dt^2+g_{rr}dr^2+r^2d\Omega_{d-2}$,
where $g_{tt}$ and
$g_{rr}$ are the time-time 
and the radius-radius components of the metric, respectively,
which in general depend on $t$ and $r$,
and
$d\Omega_{d-2}^2$ is the line element on a $d-2$ sphere,
$d\Omega_{d-2} = d\theta_1^2+\sin
\theta_1^2\,d\theta_2^2+\cdots+
\prod_{i=2}^{d-3}\sin\theta_i^2\,d\theta_{d-2}^2$.
Then, the electrovacuum Einstein-Maxwell field equations
yield that 
$g_{tt}$ and
$g_{rr}$ only depend on $r$, and indeed
the line element has the $d$-dimensional Reissner-Nordstr\"om
form given
by
\begin{equation}
ds^2 = - f\left(r\right) dt^2 + \frac{dr^2}{f\left(r\right)} + r^2
d\Omega_{d-2}\,,
\label{eq:line_element0}
\end{equation}
with the metric potential $f(r)$ being given by
\begin{equation}
f\left(r\right) = 1 - \frac{2\mu M}{r^{d-3}}
+ \frac{\chi Q^2}{r^{2\left(d-3\right)}}-{k}\,\frac{r^2}{l^2}\,,
\label{eq:line_element}
\end{equation}
where 
$M$ is the mass of the black hole, $Q$ is its
electric charge, $l^2 = \frac3{|\Lambda|}$ is the length scale related
with the absolute value of the cosmological constant $\Lambda$,
${k} = -1, 0, 1$ for spacetimes with
negative, zero or positive cosmological constant, respectively,
and $\mu$ and $\chi$ are defined as
$\mu =\frac{8\pi}{\left(d-2\right)\Omega_{d-2}}$,
$\chi =\frac{8\pi}{\left(d-2\right)\Omega_{d-2}}$,
and 
$\Omega_{d-2} = \frac{
2\pi^{\frac{d-1}{2}}}{ \Gamma\left(\frac{d-1}{2}\right)}$
is
the area of the $(d-2)$-dimensional unit
sphere with $\Gamma$ being the gamma function.
The time coordinate range is 
$-\infty<t<\infty$, the radial coordinate range is $r_+<r<\infty$,
with $r_+$ being the black hole event horizon, and 
the angular 
coordinate ranges are $0\leq\theta_i\leq\pi$ for
$i=1,...,d-3$, and 
 $0\leq\theta_{d-2}<2\pi$.
The electric potential $\varphi$ of the spacetime is 
\begin{equation}
\varphi=\frac{Q}{\left(d-3\right)r^{d-3}}\,,
\label{eq:electricpot}
\end{equation}
i.e., a Coulomb electric potential.

\subsection{The black hole horizon radius $r_+$
and the extremal black hole: $f(r)$ and its factorization}

A nonextremal black hole has two characteristic radii, the event
horizon radius $r_+$ and the Cauchy horizon radius $r_-$.
The black
hole horizon
radius $r_+$ is one of the solutions
of the equation
$f(r_+)=0$, where $f(r)$ is given in Eq.~\eqref{eq:line_element}.
Thus, in general 
$r_+=r_+\left(M,Q,l,d\right)$.
There is another possible radius 
for which
$f(r)=0$.
It is the Cauchy horizon radius $r_-$ obeying $r_-\leq r_+$,
with 
$r_-=r_-\left(M,Q,l,d\right)$.
Thus, in general one has 
\begin{equation}
r_+ =r_+\left(M,Q,l,d\right)\,,\quad\quad
r_- =r_-\left(M,Q,l,d\right)\,.
 \label{r+r-}
\end{equation}
In terms of $r_+$ and $r_-$, the
function $f(r)$ of Eq.~\eqref{eq:line_element}
can be written as
\begin{equation}
f\left(r\right) = \frac{1}{r^2} \left(r-r_+\right)
\left(r-r_-\right)g(r)\,,
\label{eq:f_factorizationnonextremal}
\end{equation}
where
$g(r)\equiv
\frac{p_{2\left(d-4+{k}^2\right)}\left(r;
{k}\right)}{r^{2\left(d-4\right)}}$,
and 
$p_{2\left(d-4+{k}^2\right)}\left(r; {k} \right)$ is a
polynomial function of $r$, of degree $2\left(d-4+{k}^2\right)$,
whose coefficients can depend on ${k}$.

An extremal black hole has one characteristic radius,
the event
horizon radius has now the same value as
the the Cauchy horizon radius $r_+=r_-$.
For the BSW mechanism, the most
interesting black hole, the one from which
large amounts of energy can be extracted
from the particle collision, is the extremal
black hole. For an extremal black hole,
there are two conditions, the former one $f(r_+)=0$,
and a new one
$\frac{df}{dr}\left(r_+\right)=0$, and they are such
the black hole horizon $r_+$ and
the Cauchy horizon $r_-$  indeed coincide,
i.e.,
\begin{equation}
r_+\left(M,Q,l,d\right)=
r_-\left(M,Q,l,d\right)\,.
 \label{r+}
\end{equation}
From the two conditions on $f(r)$
one finds that
$
r_+^{d-3}
\left(1- \frac{d-2}{d-3} {k} \frac{r_+^2}{l^2}\right)
= \mu M$.
For ${k}=0$, one
readily obtains $r_+^{d-3} = \mu M$. Using again the
two conditions on $f(r)$ one gets
that $r_+$ of an extremal black hole obeys
$r_+ =
\sqrt{
- {k} l^2R(M,Q,d) + l^2\sqrt{R^2(M,Q,d)  +
\left(\frac{d-3}{d-2}\right)^2 \left(\frac{\mu^2 M^2}{\chi Q^2}
-1\right)}
}$,
where we have used the abbreviation
$R(M,Q,d) \equiv\frac{1}{2} \frac{\mu^2 M^2}{\chi Q^2}
\left(\frac{d-3}{d-2}\right)^2 \left[\frac{2\left(d-2\right)}{d-3}
\left(1-\frac{\chi Q^2}{\mu^2 M^2}\right) - 1 \right]$.
Note that we still have
$r_+=r_+(M,Q,l,d)$, and we stick to the horizon
radius notation $r_+$, knowing that this is the extremal case 
$r_+=r_-$.
Moreover, since $r_+=r_-$, and each radius is a root of
$f(r)$, in the extremal case,
one has that $f(r)$ of Eq.~\eqref{eq:line_element} has a double route,
and so it  
assumes the form
\begin{equation}
f\left(r\right) = \frac{1}{r^2} \left(r-r_+\right)^2
g(r)\,,
 \label{eq:f_factorization}
\end{equation}
where
$g(r)\equiv
\frac{p_{2\left(d-4+{k}^2\right)}\left(r;
{k}\right)}{r^{2\left(d-4\right)}}$,
and 
$p_{2\left(d-4+{k}^2\right)}\left(r; {k} \right)$ is a
polynomial function of $r$, of degree $2\left(d-4+{k}^2\right)$,
whose coefficients can depend on ${k}$. Notice that, with
this factorization, the electric potential defined in
Eq.~\eqref{eq:electricpot} can be written as
$\varphi=\frac{Q}{\left(d-3\right)r_+^{d-3}}
\left(1-\sqrt
\frac{f(r)}{g(r)}\,
\right)^{d-3}$.

For ${k}=-1$ or ${k}=0$, in even
dimensions $g(r)$ has no real zeros, and in odd dimensions
it has one negative zero and the other zeros are not real.
For ${k}=1$, i.e., for positive cosmological constant,
there is yet another zero of $f(r)$. It is given by the cosmological
horizon radius $r_{\rm c}$, with
$r_{\rm c}=r_{\rm c}\left(M,Q,l,d\right)$
and $r_-\leq r_+\leq r_{\rm c}$,
so it is the largest horizon radius. We will be interested
in the black hole $r_+$ radius for the collision processes,
so $r_{\rm c}$ will not appear in our
developments. In this case of ${k}=1$,
in fact, one can work with the variables 
$\{M,Q,l,d\}$,  or with the
variables $\{r_+,r_-,r_{\rm c},d\}$, or with a
combination of the two.
The most natural combination of variables
in the problem at hand
is $\{r_+,Q,l,d\}$.
In the case of ${k}=0$ and ${k}=1$ there is 
no cosmological
horizon and one can work with the
natural combination of variables
in the problem at hand, i.e., $\{r_+,Q,l,d\}$.

\subsection{Equations of motion for particles and the electric
ergosphere}

\subsubsection{Equations of motion for particles}

Neutral particles follow geodesics
in a Reissner-Nodtsr\"om spacetime, but electrically
charged particles do not.
To obtain 
the equations of motion for a charged particle
it is useful to resort to the Lagrangian of
the particle and its Euler-Lagrange equations of motion. 
Consider a particle with mass $m$ and
specific charge $\tilde e$,
moving in a $d$-dimensional Reissner-Nodtsr\"om spacetime.
The Lagrangian $L$ for the particle can then be written
as
$L =\frac12 g_{ab}u^au^b-A_au^a$,
where $u^a=\frac{dx^a}{d\tau}$ is the four-velocity
of the particle,
$\tau$ is its proper time defined through $d\tau^2=-ds^2$,
and 
$A_a$ is the electromagnetic
four-potential.
For the
Coulomb interaction of the particle with the black hole
spacetime one has $A_a=\varphi\delta^t_a$,
and assuming pure radial motion, i.e., 
$u^a={u^t}\delta^a_t+{u^r}\delta^a_r$,
the Lagrangian $L$ for the particle is then
given by
$L =\frac12\left( - f\left(r\right) \dot{t}^2 +
\frac{\dot{r}^2}{f\left(r\right)} -
\frac{2 {\tilde e} Q}{\left(d-3\right) r^{d-3}} \dot{t}\right)$,
where a dot means a derivative relative to the
particle's
proper time $\tau$, and we have used $g_{tt}=-f$,
$g_{rr}=\frac1f$, $\varphi=\frac{Q}{\left(d-3\right)r^{d-3}}$,
$u^t=\dot t$, and $u^r=\dot r$.
From the Euler-Lagrange equations for $L$ one obtains
that the equation of motion for
the  coordinate $t$ is $\frac{d \;}{d
\tau}\left(\frac{\partial L}{\partial\dot t}\right)=0$, i.e.,
$\frac{d \;}{d \tau}\left(f{\dot t}+\frac{ {\tilde e}
Q}{\left(d-3\right) r^{d-3}} \right)=0$, which
can be integrated to 
 ${\dot t}=\frac{E-\frac{
e Q}{\left(d-3\right) r^{d-3}}}{mf}$, 
where $E$ is the energy of the
particle, a conserved quantity, and 
the particle's total electric charge has been defined
as $e=m\tilde e$.
For pure radial motion, one
finds that the first integral
for the radial coordinate is 
$\dot{r}^2 = \frac{\left( E-\frac{ e Q}{\left(d-3\right)
r^{d-3}}\right)^2}{m^2}-f$.  Defining the
four-momentum as $p^a\equiv m u^a$,
the time component of the
four-momentum of the particle is $p^t\equiv m \dot{t}$ and the
radial  component 
is $p^r\equiv m \dot{r}$. One then finds that the
equations of motion for the
particle can be put in the form
\begin{equation}
p^t\equiv m \dot{t} = \frac{X}{f},
\quad\quad\quad\quad
p^r \equiv m \dot{r} = \varepsilon Z\,,
\label{eq:motion_t_penrose}
\end{equation}
where the quantities $X$ and $Z$ are defined as
\begin{equation}
X(r) = E - \frac{ e Q}{\left(d-3\right) r^{d-3}},
\quad\quad\quad\quad
Z(r)  = \sqrt{X^2 - m^2 f(r)},
  \label{eq:motion_r_penrose}
\end{equation}
with $f=f(r)$ being given in Eq.~\eqref{eq:f_factorization},
and $\varepsilon = \pm 1$ defining the direction of
the particle's motion,
$-1$ for inward radial  motion and $+1$
for outward motion.
The forward in time condition, $\dot{t} > 0$
implies that $X > 0$ outside the horizon. In what follows,
it will always be assumed
that the electric charge of the black
hole is positive, $Q > 0$, without loss of generality.

\subsubsection{The electric ergosphere}

Given the forward in time condition, $\dot{t} > 0$,
one has $X > 0$ for any radii outside the horizon. 
From Eq.~\eqref{eq:motion_r_penrose}
this implies $E -\frac{ e Q}{\left(d-3\right) r^{d-3}}>0$.
Given a black hole electric charge $Q$, with $Q$ positive
as assumed, it is clear
from the latter
expression 
that one can have negative energy states $E$ for the particle,
as long as its electric charge $e$ is negative
and the position of
the particle $r$ is sufficiently small.
Writing for these negative energy states $E=-\lvert E \rvert$
and $e=-\lvert e \rvert$, we have
that indeed these states exist when the
radial position $r$ of the particle obeys
\begin{equation}
r_+\leq r<
r_{\mathrm{ergo}}\,,\quad\quad\quad\quad\quad\quad
r_{\mathrm{ergo}}^{d-3} =\frac{\lvert e \rvert Q}{\left(d-3\right)
\lvert E \rvert }\,,\quad\quad\quad
E<0\,,\;\; e<0.
\label{eq:ergosphere}
\end{equation}
The region $r_+\leq r<
r_{\mathrm{ergo}}$ is an 
ergosphere that comes from the existence of electric
charge, and is called 
electric ergosphere or generalized
ergosphere.

\section{
Definitions and necessary conditions for energy production from BSW
particle collisions: Definition of critical particles and analysis of
the energy at the center of mass frame
}
\label{sec:ECM}

\subsection{Definition of critical particles
and energy at the center of mass frame}

\subsubsection{Critical, near-critical, and usual particles}

We nominate each particle as particle $i$, in general
$i=1,2,3,4$. Each particle has attributes like
its energy $E_i$, its mass $m_i$, its electric charge $e_i$,
and so on.
In what follows, definitions of  critical, near-critical, and usual
particles will be important. Thus, we establish here 
definitions for thee different particles, see \cite{zaslavskii_2010}.

The horizon radius $r_+$ 
obeys the equation
$f(r_+)=0$ as we have seen. If in addition $X(r_+)$ in
Eq.~\eqref{eq:motion_r_penrose}
is also zero, $X(r_+)=0$,
then $p^t$ is undetermined a priori and
interesting things can happen.
When
for a given particle $i$ one has
$X_i(r_+)=0$, then
for a value $E_i$ of the energy of the particle there
corresponds a definite value of the electric charge,
the critical electric charge ${e_i}_c$  given by
\begin{equation}
{e_i}_c = \frac{r_+^{d-3} \left(d-3\right)}{Q}\,E_i.
\label{eq:critical_charge_definition}
\end{equation}
A particle $i$
is then
defined as critical if its electric charge $e_i$ is equal to the
critical electric charge, i.e.,
\begin{align}
e_i= {e_i}_c \,,\quad\quad {\rm critical\; particle}\,.
\label{eq:criticalparticle}
\end{align}
A particle $i$
is defined as near critical if its electric charge $e_i$ is
almost equal to the critical
electric charge, i.e.,
\begin{align}
e_i= {e_i}_c
\left(1 +
\delta\right)\,,\quad\quad {\rm near\; critical\; particle}\,, 
\label{eq:nearcriticalparticle}
\end{align}
with $|\delta| \ll 1$ and $\delta$
positive or negative. If $\delta=0$,  one
recovers the definition of a critical particle.
A particle $i$
is defined as usual if its electric charge $e_i$ differs
from the critical and near-critical
electric charges, i.e.,
\begin{align}
e_i\neq {e_i}_c \left(1 +
\delta\right)\,,\quad\quad {\rm usual\; particle}\,,
\label{eq:usualparticle}
\end{align}
i.e., the electric charge of the particle is
significantly different from the critical electric charge.
These definitions are important.

\subsubsection{Particle collision and 
energy at the center of mass frame}

To study the energy generated from the BSW effect, a collision
between two ingoing
particles is assumed to occur in a $d$ dimensional
extremal Reissner-Nordstr\"om black hole
spacetime with
horizon radius $r_+$, electric charge $Q$, and 
cosmological constant ${k}
\Lambda$.
The mass
of particle $i$
is denoted as $m_i$, and the
electric charge of particle $i$
is denoted as $e_i$, so that, before the collision, 
particles $i= 1, 2$
have masses and charges
$m_1$ and $e_1$, and $m_2$ and $e_2$, respectively,
and after the collision, 
particles $i= 3,4$
have masses and charges
$m_3$ and $e_3$, and $m_4$ and $e_4$, respectively,
assuming, as we do, that two particles come out
of the collision.

To have a grasp on the collision process and to understand
the
necessary conditions for energy extraction, we
start by making a simplifying assumption,
the incoming masses are equal 
$m_1 = m_2$, although the electric charges $e_1$ and $e_2$
are different. So we put $m\equiv m_1 = m_2$.
An important quantity is the energy
of the center of mass $E_{\mathrm{CM}}$. 
To calculate an expression for it, note that
the total four-momentum vector
$p^a$
is 
$p^a=p_1^a+p_2^a$, where $p_1^a$ is the
momentum of particle 1 and $p_2^a$ 
 is the
momentum of particle 2.
In the center of mass the total three-momentum
is zero, so the four-momentum is
only composed of the center of mass energy
in the local frame
and we can write in the center of mass
${\bar p}^l=E_{\rm CM}\delta_t^l$.
But $p^a={\bar p}^le^a_l$ where 
$e^a_l$ is the local tetrad.
So, $p^2=g_{ab}p^ap^b=
g_{ab}{\bar p}^l{\bar p}^me^a_le^b_m=
{\bar p}^l{\bar p}^m
\eta_{lm}=E_{\rm CM}^2\delta_t^l\delta_t^m
\eta_{lm}=E_{\rm CM}^2\eta_{tt}=-E_{\rm CM}^2
$, where $g_{ab}$ is here the Reissner-Nordstr\"om metric and 
$\eta_{lm}$ is the Minkowski metric.
On the other hand
$p^2=g_{ab}p^ap^b=g_{ab}(p^a_1
+p^a_2
)
(p^b_1
+p^b_2
)
=g_{ab}p_1^ap_1^b
+g_{ab}p_2^ap_2^b
+
2g_{ab}p_1^ap_2^b
$. Now $p_1^a=mu_1^a$ and $p_2^a=mu_2^a$,
so $g_{ab}p_1^ap_1^b=-m^2$,
$g_{ab}p_2^ap_2^b=-m^2$, and 
$2g_{ab}p_1^ap_2^b=
2m^2g_{ab}u_1^au_2^b$.
Since at $r$,
$p^2$ is an invariant, one gets
$-(E_{\mathrm{CM}}^2)=-2m^2+2m^2g_{ab}u_1^au_2^b$, i.e.,
$\frac{E_{\mathrm{CM}}^2}{2m^2} = 1 - g_{ab}u_1^au_2^b$.
For a pure radial collision one has
$u_1^a={u_1^t}\delta^a_t+{u_1^r}\delta^a_r$
and
$u_2^a={u_2^t}\delta^a_t+{u_2^r}\delta^a_r$,
so that
$\frac{E_{\mathrm{CM}}^2}{2m^2} = 1 - g_{tt}u_1^t u_2^{t}
- g_{rr}u_1^r u_2^r
$.
Since 
$u_1^t={\dot t_1}$,
$u_1^r={\dot r_1}$,
$u_2^t={\dot t_2}$,
$u_2^r={\dot r_2}$,
 $g_{tt}=-f$, and
$g_{rr}=\frac1f$,
one finds
$\frac{E_{\mathrm{CM}}^2}{2m^2} = 1 + f {\dot t_1}{\dot t_2}
- \frac{ {\dot r_1}{\dot r_2}}{f}$.
Using the equations of motion given in 
Eq.~\eqref{eq:motion_t_penrose}, one finds that
\begin{equation}
\frac{E_{\mathrm{CM}}^2(r)}{2m^2} =
1 + \frac{X_1(r) X_2(r) - Z_1(r) Z_2(r)}{m^2f(r)},
\label{eq:ECM2}
\end{equation}
where $X_i$ and $Z_i$, $i=1,2$, are the quantities defined in
Eq.~\eqref{eq:motion_r_penrose} evaluated for particle $i=1, 2$ at
the point of collision $r$,
the value $\varepsilon_i=-1$, $i=1,2$,
was used since we assume a BSW collision and so
the two particles move inwards, 
and $f$ is given in
Eq.~\eqref{eq:f_factorization}.  $E_{\rm CM}$ which clearly is a
function of $r$, $E_{\rm CM}(r)$, is to be computed at some
radius, possibly near the black hole horizon $r_+$.
Note that after the collision,
particles 3 and 4 have the same $E_{\rm CM}$ as before the
collision.

Three types of collision between the incoming particles can happen.
They are the  collision between two critical particles, the
collision between a critical particle and a usual particle,
and the collision between a near-critical particle and a
usual particle. For each type of collision, the energy at the center
of mass frame can be computed and the conditions which particles must
obey to allow the occurrence of the collision at the horizon can be
established.

\subsection{The three types of collisions: Estimates
for the produced energy}

\subsubsection{Collision between two critical particles}

Here we study
a collision between two ingoing critical particles in the
vicinity of the extremal black hole event horizon, at some radius $r$
which is near or at $r_+$. 
So we assume
that particle 1 is exactly critical and particle 2 is
exactly critical. Then particle 1 has $e_1= {e_1}_c$,
see Eq.~\eqref{eq:criticalparticle},
and
from Eq.~\eqref{eq:motion_r_penrose} we have
$X_1 = E_1 \left(1 - \left(\frac{r_+}{r}\right)^{d-3}\right)$
and
$Z_1 = \sqrt{E_1^2 \left(1 -
\left(\frac{r_+}{r}\right)^{d-3}\right)^2 - m^2 f}$.
Particle 1 can only move for radii for which $Z_1$ is well
defined, in the sense that the argument of the square root has to be
positive, $E_1^2 \left(1 - \left(\frac{r_+}{r}\right)^{d-3}\right)^2 -
m^2 f > 0$. Therefore, particle 1 can only reach the horizon provided
that the term $-m^2 f\left(r\right)$ approaches zero faster than
$E_1^2 \left(1 - \left(\frac{r_+}{r}\right)^{d-3}\right)^2$ when
$r\to r_+$.
Particle 2 is critical, it has $e_2= {e_2}_c$,
see Eq.~\eqref{eq:criticalparticle},
and
from Eq.~\eqref{eq:motion_r_penrose} we have
$X_2 = E_2 \left(1 - \left(\frac{r_+}{r}\right)^{d-3}\right)$
and
$Z_2 = \sqrt{E_2^2 \left(1 -
\left(\frac{r_+}{r}\right)^{d-3}\right)^2 - m^2 f}$.
Particle 2 can as well only move for radii for which $Z_2$ is well
defined, in the sense that the argument of the square root has to be
positive, $E_2^2 \left(1 - \left(\frac{r_+}{r}\right)^{d-3}\right)^2 -
m^2 f > 0$. Therefore, particle 2 can only reach the horizon provided
that the term $-m^2 f\left(r\right)$ approaches zero faster than
$E_2^2 \left(1 - \left(\frac{r_+}{r}\right)^{d-3}\right)^2$ when
$r\to r_+$.
To find if particles 1 and 2 can reach the black hole
event horizon, one has
to study how the metric potential $f\left(r\right)$ approaches zero
when $r\to r_+$. This will depend on the factorization of this
function $f\left(r\right)$.
The case in which $f$ goes faster to zero when the horizon $r_+$
is approached is the extremal case with
the factorization obtained in Eq.~\eqref{eq:f_factorization},
since $1 - \frac{r_+}{r}\leq
1 - \left(\frac{r_+}{r}\right)^{d-3}$. The factorization for
nonextremal black holes,
Eq.~\eqref{eq:f_factorizationnonextremal}, shows that
the approach is not fast enough.
The generic expression for $Z_1$ in the extremal case is then
$Z_1 = \left(1-\left(\frac{r_+}{r}\right)^{d-3}\right)
\sqrt{E_1^2-m^2\, g(r) \,
\frac{\left(1-\frac{r_+}{r}\right)^2}{\left(1-
\left(\frac{r_+}{r}\right)^{d-3}\right)^2}}$.
Since $\lim_{r\to r_+}
\frac{\left(1-\frac{r_+}{r}\right)^2}{\left(1-
\left(\frac{r_+}{r}\right)^{d-3}\right)^2}
= \frac{1}{\left(d-3\right)^2}$,
$Z_1$ near the horizon is given by
$
Z_1 = \left(1-\left(\frac{r_+}{r}\right)^{d-3}\right) \sqrt{E_1^2-
\frac{m^2}{\left(d-3\right)^2}\,
g(r) }
$.
Note that, by definition, the function
$g(r)$
is always
nonzero at the horizon, since the part of $f\left(r\right)$ that
vanishes at the horizon was already factorized. Therefore, $Z_1$ is
well defined and the horizon is reachable by a critical particle
provided that $E_1^2 >
\frac{m^2}{\left(d-3\right)^2}\,
g(r_+)$. The same can be said for $Z_2$,
so that 
$
Z_2 = \left(1-\left(\frac{r_+}{r}\right)^{d-3}\right) \sqrt{E_2^2-
\frac{m^2}{\left(d-3\right)^2}\,
g(r) }
$, therefore, $Z_2$ is
well defined and the horizon is reachable by a critical particle
provided that $E_2^2 >
\frac{m^2}{\left(d-3\right)^2}\,
g(r_+)$.
Thus in brief, we can write
\begin{equation}
\begin{aligned}
&X_1(r) = E_1 \left(1 - \left(\frac{r_+}{r}\right)^{d-3}\right),
\quad\quad
Z_1 = \left(1-\left(\frac{r_+}{r}\right)^{d-3}\right) \sqrt{E_1^2-
\frac{m^2}{\left(d-3\right)^2}\,
g(r) }\,,
\\
&
X_2(r) = E_2 \left(1 - \left(\frac{r_+}{r}\right)^{d-3}\right),
\quad\quad
Z_2 = \left(1-\left(\frac{r_+}{r}\right)^{d-3}\right) \sqrt{E_2^2-
\frac{m^2}{\left(d-3\right)^2}\,
g(r) }
\,,
\end{aligned}
\label{eq:XZcriticalcritical}
\end{equation}

Substituting now $X_1$, $Z_1$,
$X_2$, and $Z_2$ 
of Eq.~\eqref{eq:XZcriticalcritical}
in the expression for the energy
at the center of mass frame for a collision,
Eq.~\eqref{eq:ECM2},
when the collision is between a critical and
another critical particle at the extremal horizon $r_+$,
one gets
\begin{equation}
\frac{E_{\mathrm{CM}}^2(r)}{2m^2} = 1 + \frac{\left(d-3\right)^2}
  {m^2g(r_+)}
  \left[E_1  E_2 - \sqrt{E_1^2 -
\frac{m^2}{\left(d-3\right)^2}
  g(r_+)}
\sqrt{E_2^2 -
\frac{m^2}{\left(d-3\right)^2}
  g(r_+)}\;\right].
\label{eq:ECM6}
\end{equation}
Note that, as displayed in Eq.~\eqref{eq:ECM6},
$E_{\mathrm{CM}}(r)$ does not depend on $r$
in zero order in $r-r_+$. So, $E_{\mathrm{CM}}$
remains finite at the horizon,
\begin{align}
E_{\mathrm{CM}}(r_+) =  {\rm finite}\,,\quad\quad
{\rm particle\,1\,critical}\,\,e_1= {e_1}_c,
\quad
{\rm particle\,2\,critical}\,\,e_2= {e_2}_c\,.
\label{eq:limit01}
\end{align}
Therefore, there is no great gain. When two critical particles collide
near or at the horizon there is no great amount or even no
energy generation.

\subsubsection{Collision between a critical and a usual particle}

Here we study
a collision between an ingoing critical particle
and an ingoing usual particle in the
vicinity of the extremal black hole event horizon, at some radius $r$
which is near or at $r_+$.
So, we assume
that particle 1 is exactly critical and particle 2 is
usual. Then particle 1 has $e_1= {e_1}_c$,
see Eq.~\eqref{eq:criticalparticle},
and
from Eq.~\eqref{eq:motion_r_penrose} we have
$X_1 = E_1 \left(1 - \left(\frac{r_+}{r}\right)^{d-3}\right)$
and
$Z_1 = \sqrt{E_1^2 \left(1 -
\left(\frac{r_+}{r}\right)^{d-3}\right)^2 - m^2 f}$.
Since particle 1 is critical, the calculations
previously made hold here, and
we do not repeat them, noting that $f$ has to be
the extremal black hole function in order to have
a possible gain.
Particle 2 is usual,
it has
$e_2\neq {e_2}_c \left(1 + \delta\right)$, see
Eq.~\eqref{eq:usualparticle},
and one simply has that
$
X_2 = E_2 - \frac{ e_2 Q}{\left(d-3\right) r^{d-3}}$ and 
$Z_2 = \sqrt{\left(E_2 - \frac{ e_2
Q}{\left(d-3\right) r^{d-3}}\right)^2 - m^2 f}$.
To find if particle 2 can reach the black hole event horizon, one has
to study how the metric potential $f\left(r\right)$ approaches zero
when $r\to r_+$.
Due to the forward in time condition,
one has that $X_2 > 0$ for all possible
values of the radial coordinate. This means that, when $r \to r_+$,
the term $\left(E_2 - \frac{ e_2 Q}{\left(d-3\right)
r^{d-3}}\right)^2$ remains positive, while the term $-m^2
f\left(r\right) \to 0$. Therefore, particle 2 can always reach the
horizon.
Thus in brief, we can write
\begin{equation}
\begin{aligned}
&X_1(r) = E_1 \left(1 - \left(\frac{r_+}{r}\right)^{d-3}\right),
\quad\quad
Z_1 = \left(1-\left(\frac{r_+}{r}\right)^{d-3}\right) \sqrt{E_1^2-
\frac{m^2}{\left(d-3\right)^2}}\,,
\\
& X_2(r) = E_2 - \frac{ e_2 Q}{\left(d-3\right) r^{d-3}},
\quad\quad\;\;\; Z_2(r) = \sqrt{\left(E_2 - \frac{ e_2
Q}{\left(d-3\right) r^{d-3}}\right)^2 - m^2 f(r)}\,,
\end{aligned}
\label{eq:Z1non_extremal}
\end{equation}

Substituting now $X_1$, $Z_1$,
$X_2$, and $Z_2$ 
of Eq.~\eqref{eq:Z1non_extremal}
in the expression for the energy
at the center of mass frame for a collision,
Eq.~\eqref{eq:ECM2},
when the collision is  between a critical and a
usual particle one gets
the expression,
\begin{equation}
\frac{E_{\mathrm{CM}}^2(r)}{2m^2} = 1 +
\frac{
E_2 - \frac{ e_2 Q}{\left(d-3\right) r_+^{d-3}}
}{m^2
g(r_+)}
\left[E_1 -
\sqrt{
E_1^2 -
\frac{m^2}{\left(d-3\right)^2}\,
g(r_+)
}
\; \right]
\frac{
d-3
}
{1-\frac{r_+}{r}}\,,
\label{eq:ECM5}
\end{equation}
for a collision happening at a radius $r$ near $r_+$.
When $r \to r_+$, the factor
$\frac{1}{1-\frac{r_+}{r}}
\to \infty$. Thus, the energy at the center of mass frame explicitly
diverges
when the collision is at
the horizon. Therefore, a collision between a critical and
a usual particle in the vicinity of the event horizon leads to a
divergent energy at the center of mass frame
\begin{align}
E_{\mathrm{CM}}(r_+) = \infty\,,\quad\quad
{\rm particle\,1\,critical}\,\,e_1= {e_1}_c,
\quad{\rm particle\,2\,usual}\,.
\label{eq:limit2}
\end{align}
Therefore, here one can have a great gain.
 When two particles collide
near or at the horizon, one of them is critical and the other is usual,
there is the possibility of a great amount of
energy generation.

\subsubsection{Collision between a near-critical and a usual particle}
We
now consider a collision between an ingoing
near-critical particle, which
for definiteness is chosen to be particle 1, and an
ingoing usual particle,
which is particle 2. The energy at the center
of mass frame for collisions happening at the black
hole event horizon, Eq.~\eqref{eq:ECM2}, is exactly
$
\frac{E_{\mathrm{CM}}^2}{2m^2} = 1 + \frac{1}{2}
\left(\frac{{e_2}_c - e_2}{{e_1}_c  - e_1} +
\frac{{e_1}_c  -  e_1}{{e_2}_c - e_2}\right)
$.
Now, particle 1 being near critical has an
electric charge given by $e_1 = {e_1}_c
\left(1 + \delta\right)$, with $\lvert \delta  \rvert \ll 1$,
and $\delta$ has to be negative or zero, $\delta\leq0$,
since to arrive 
at the horizon the electric charge
of the particle has to be less or equal
to the critical charge due to the forward in
time condition, i.e., $e_1 \leq {e_1}_c$. So, 
$\frac{{e_1}-{e_1}_c}{{e_1}_c}=\delta$.
Thus, the energy at center of
mass frame scales as the inverse of the square root of $|\delta|$.
Indeed,
\begin{equation}
\frac{E_{\mathrm{CM}}^2(r)}{2m^2} =  \frac{1}{2}
\frac{{e_2}_c-{e_2} }{
{e_1}_c  - {e_1}_c
}
+ 1\,,
\label{eq:ECM4}
\end{equation}
plus
$\mathcal{O}\left(
\frac{{e_1} -{e_1}_c}{{e_1}_c}\right)=\mathcal{O}\left(\delta\right)$
for a collision at $r_+$.
Therefore, since the denominator is
equal to ${e_1}_c|\delta|$,
the energy at the center of mass frame can be made as large
as one wants. In particular, if $\delta=0$, i.e., if particle 1 is
exactly critical, the energy seems to diverge. However, to know
whether this divergence is possible or not, one has to assume from the
very beginning that particle 1 is exactly critical, in order to
find
how the horizon can be reached by such a particle. This is what we
have done in the previous case, and no need to take this case into account.

\subsection{Picking
the only available interesting type of collision}

The energy at the center of mass frame was computed for three
different types of collisions between two
ingoing particles with equal masses
and different electric charges. The types of collision are between two
critical particles, between a critical and a usual particle, and
between a near-critical and a usual particle.  For the first type, a
collision between two critical particles, it was verified whether these
particles were able to reach the black hole event horizon, and it was
found that this is only the case if one assumed an extremal black
hole, and it was further
found that there is no divergence in the energy at
the center of mass frame.  For the second type,
a collision between a critical particle and a usual particle,
 it was verified whether these
particles were able to reach the black hole event horizon,
and it was
found that this is only the case if one assumed an extremal black
hole, and it was further
found that there is divergence in the energy at
the center of mass frame when the collision occurs exactly at the
black hole event horizon.  For the third type,
a collision between a near critical particle and a usual particle,
it was concluded that
the energy at the center of mass frame can be as large as one
wants. However, this quantity is always finite, diverging only in the
limit in which particle 1 is exactly critical, so
one is back in second type.

In brief, the only interesting type
of collision, the one that yields a divergingly
high energy generation, is the
second one, i.e., a collision between a critical and a usual particle
in an extremal black hole background.
However, even when the energy at the center of mass frame is
divergingly high, it can happen that energy extraction is not
possible, for instance, the particles that come out
of the collision all enter into the black hole. 
Thus, one is advised to analyze carefully the collision process
and consider the general case
in which the particles have different masses.  Therefore, for the
situations in which the energy at the center of mass frame is
divergent, we analyze a collision between
ingoing particles $i = 1, 2$
with masses $m_1$ and $m_2$ that can be different in general,
originating two final particles, $i = 3, 4$, with masses $m_3$
and $m_4$,
also
different in general. For definiteness, it is assumed that particle 3 is
outgoing in the end of the process,
while particle 4 is ingoing and
falls into the black hole. Thus,
the energy extracted essentially
equals the energy of the emitted particle, in
this case the energy of particle 3. We want
to establish bounds on the
energy of the emitted particle, finding under which conditions energy
extraction through a Penrose
process is possible using the BSW effect.  Thus, we aim to
get a better
measure, better than simply unbounded energy
at the center of mass frame, of extracted energy
from the black hole due to
the BSW effect when combined with
the Penrose process in the collisional Penrose
process.

\section{Penrose and super-Penrose
energy extraction from a collision between a critical and a
usual particle: Full story }
\label{sec:E_emitted}

\subsection{Conservation laws and energy of critical and usual particles}

Up to now, it has been shown that, under certain conditions,
a BSW collision between two particles in the vicinity of the horizon
of an extremal black hole can
lead to a divergence of the energy at the center of mass frame. This
happens for a collision between particle 1,
a critical particle, and particle 2, a
usual particle. However, this divergence is not a
sufficient condition to guarantee that there is some outgoing
particle  with some net energy, let alone an unbounded 
energy.  In this section, we find the energy and the mass of a
particle which results from a collision near the black horizon
and is sent to larger radii, i.e.,
we examine the collisional Penrose process
in which the collisions are of BSW type.
For
spacetimes with zero or positive cosmological constant,
i.e., asymptotically flat or asymptotically dS spacetimes, 
the escaping
particle can reach infinity,  for
spacetimes with negative cosmological constant,
i.e., asymptotically AdS spacetimes, 
this escaping
particle only reaches infinity if it is massless.

We assume a collision between the initial
ingoing particles $i = 1, 2$ and now consider the more general
case in which the particles can have different masses, i.e., particle 1
has mass $m_1$ and
electric charge $e_1$ and particle 2 has mass $m_2$
and electric charge $e_2$.
We further assume that two particles $i = 3, 4$ issue
after the collision,
with mass $m_3$ and electric charge $e_3$,
and mass $m_4$ and electric charge $e_4$.
We consider that 
energy, radial momentum, and electric charge
are conserved in the collision process,
which implies the following three conservation laws,
\begin{equation}
   X_1 + X_2 = X_3 + X_4,
\label{eq:energy_conservation}
\end{equation}
\begin{equation}
   \varepsilon_1 Z_1 + \varepsilon_2 Z_2 = \varepsilon_3 Z_3 +
   \varepsilon_4 Z_4,
\label{eq:momentum_conservation}
\end{equation}
\begin{equation}
   e_1 + e_2 = e_3 + e_4,
\label{eq:charge_conservation}
\end{equation}
respectively.

\begin{figure}[h]
    \centering
    \includegraphics[scale=0.4]{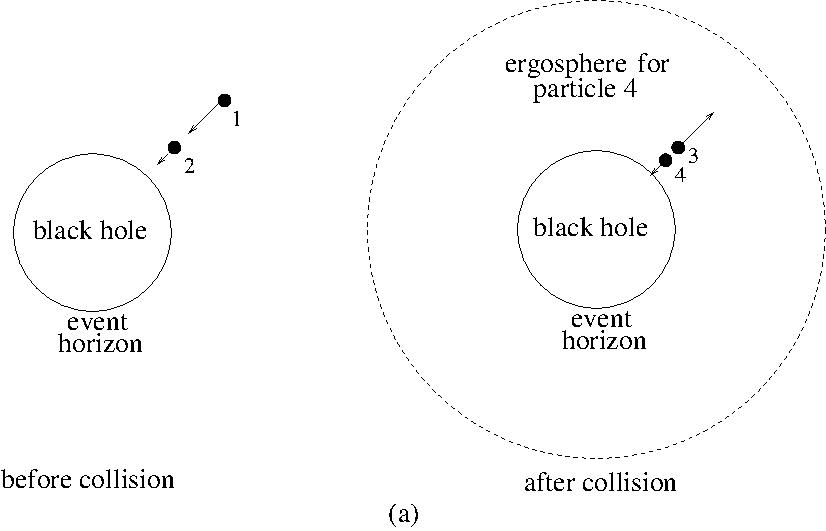}
\vskip 1.0cm
\centering
    \includegraphics[scale=0.4]{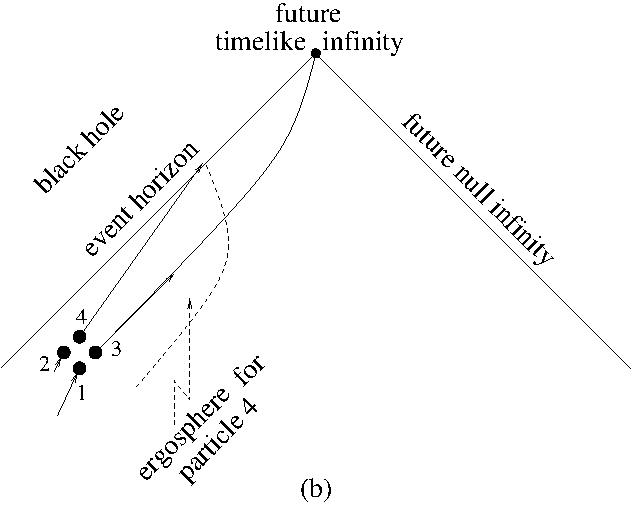}
\caption{
(a) Schematic picture for an electrically charged BSW collision. The
electrically charged particle 1 collides with the
electrically charged particle 2 outside the event horizon of
a black hole at the center. Particle 3
emerges from the collision escaping to infinity
taking with it energy in a
Penrose process, while particle 4,
which is inside its own electric ergosphere,
falls through the event horizon.  On the left a snapshot
before the collision. On the right a snapshot after the collision.
Note that the electric ergosphere only materializes after particle
4 with negative energy
is created in the collision.
(b) Carter-Penrose diagram for the collisional Penrose process in a
spacetime with zero cosmological constant.  After particles 1 and 2
suffer a BSW collision, particles 3 and 4 emerge, with the former
escaping to future timelike infinity through a Penrose process and the
latter, being in its own electric ergosphere, falling into the
black hole. Particle 3 can carry arbitrarily large energies to
infinity characterizing a super-Penrose process.
}
\label{collisionfigure}
\end{figure}

We further assume
that
particle 1 is critical and goes into the black hole
so that $\varepsilon_1=-1$,
particle 2 is usual and also goes into the black hole
so that $\varepsilon_2=-1$,
and particle 4 which is one of the particles
that comes out of the collision
is usual and goes into the black hole
so that $\varepsilon_4=-1$. The black hole
is
extremal since we have seen
that extremal black holes are the ones that can yield
large amounts of energy.
We want to find the properties of
particle 3
that emerges from the collision and is outgoing,
see Fig.~\ref{collisionfigure}.
For critical particles, the
quantity $X$ evaluated at the horizon, $X(r_+)$, vanishes,
while for usual particles one has that $X(r_+)\neq 0$.
Therefore, for particle 1,
$X_1(r_+)=0$ and its energy can be related with its
electric charge as
$E_1 = \frac{e_1 Q}{\left(d-3\right) r_+^{d-3}}$,
where Eq.~\eqref{eq:motion_r_penrose} was used.  For particle 2, which
is usual, one uses the forward in time condition, $\dot{t}>0$, to find
a lower bound for the energy.
From Eqs.~\eqref{eq:motion_t_penrose} and
\eqref{eq:motion_r_penrose}
this condition implies that
$X_2(r_+) > 0$ and, therefore,
$E_2 > \frac{e_2 Q}{\left(d-3\right) r_+^{d-3}}$
and the same applies for particle 4, 
$E_4 > \frac{e_4 Q}{\left(d-3\right) r_+^{d-3}}$.
In brief, we can write for particles 1, 2 , and 4 that
$X_1 = E_1 \left(1 - \left(\frac{r_+}{r}\right)^{d-3}\right)$,
$Z_1 = \sqrt{E_1^2 \left(1 -
\left(\frac{r_+}{r}\right)^{d-3}\right)^2 - m^2 f}$,
$X_2 = E_2 - \frac{ e_2 Q}{\left(d-3\right) r^{d-3}}$,
$Z_2 = \sqrt{\left(E_2 - \frac{ e_2
Q}{\left(d-3\right) r^{d-3}}\right)^2 - m^2 f}$,
$X_4 = E_4 - \frac{ e_4 Q}{\left(d-3\right) r^{d-3}}$,
and 
$Z_4 = \sqrt{\left(E_4 - \frac{ e_4
Q}{\left(d-3\right) r^{d-3}}\right)^2 - m^2 f}
$.
Since particle 1 is assumed critical, and therefore 
 $X_1(r_+)
= 0$ and so 
$e_1= \frac{ \left(d-3\right) r_+^{d-3}} {Q}E_1$, we can evaluate
$X(r)$ and $Z(r)$
of Eq.~\eqref{eq:motion_r_penrose} near $r_+$,
i.e., expand it in $r-r_+$, or what amounts to the same thing
in $\sqrt{f(r)}$.
Doing the expansion, one finds 
$
X_1(r)=E_1\sqrt{f(r)}\frac{d-3}{\sqrt{g(r)}}
$
and
$ Z_1(r)=E_1\sqrt{f(r)}
\sqrt{
\frac{(d-3)^2}{g(r)}
-\frac{m_1^2}{E_1^2}
}$,
plus $\mathcal{O}\left(f\right)$ in both equations.
Particle 2 and particle 4 are assumed usual,
i.e., $X_2(r_+)
\neq 0$ and $X_4(r_+)
\neq 0$, and expanding these functions near $r_+$, i.e.,
in $\sqrt{f(r)}$ one finds
$X_2(r) = X_2(r_+) + \frac{e_2\, Q}{r_+^{d-3}}
  \sqrt{\frac{f(r)}{g(r)}}$,
$Z_2(r) = X_2 (r)$,
$X_4(r) = X_4(r_+) + \frac{e_4\, Q}{r_+^{d-3}}
  \sqrt{\frac{f(r)}{g(r)}}$,
$Z_4(r) = X_4 (r)$,
plus $\mathcal{O}\left(f\right)$ in the four equations.
Collecting these results gives
\begin{equation}
\begin{aligned}
&X_1(r)=E_1\frac{d-3}{\sqrt{g(r)}}\sqrt{f(r)},
\quad\quad\quad\quad\quad\quad\quad\quad\;
Z_1(r)=E_1
\sqrt{
\frac{(d-3)^2}{g(r)}
-\frac{m_1^2}{E_1^2}
}\,\sqrt{f(r)}\,,
\\
& X_2(r)= X_2(r_+) + \frac{e_2\, Q}{r_+^{d-3}  \sqrt{g(r)},       }
  \sqrt{f(r)},
\quad\quad\;\;\; Z_2(r) = X_2 (r)\,,
\\
&
 X_4(r)= X_4(r_+) + \frac{e_4\, Q}{r_+^{d-3}  \sqrt{g(r)},       }
  \sqrt{f(r)},
\quad\quad\;\;\;  Z_4(r) = X_4 (r)\,,
\end{aligned}
\label{eq:124extremal}
\end{equation}
plus $\mathcal{O}\left(f\right)$ in all equations, and
where $X_2(r_+) = E_2 -
\frac{e_2}{\left(d-3\right)}
\frac{Q}{r_+^{d-3}}$ and
$X_4(r_+) = E_4 -
\frac{e_4}{\left(d-3\right)}
\frac{Q}{r_+^{d-3}}$.

Thus, let us summarize what we have up to now.
The situation in analysis in this section corresponds to a collision
at some radius $r$ near the
extremal horizon radius $r_+$
between a
critical and a
usual particle, since this is the case for
which the energy at the center of mass frame diverges. Therefore,
particle 1 is critical and particle 2 is usual.
We also assume that particle 4 that comes out of the collision is
usual. 
Moreover, since we are dealing
with the BSW effect, we assume
that particle 1 goes in,
particle 2 goes in,
and further assume
that particle 4 goes in.
We consider that the other final particle,
particle 3, escapes somehow after the
collision. Several situations can happen, namely,
particle 3 goes immediately out so that
at the collision  one has
$\varepsilon_3=+1$, or
particle 3 goes in first, so that at the collision 
one has $\varepsilon_3=-1$, and then
reverses direction and goes out,
or particle 3 goes in always
and $\varepsilon_3=-1$.
Now we calculate the properties of the final 
outgoing emitted particle, particle 3.

\subsection{Energy of the outgoing emitted particle}

Particle 3, the one emitted, needs a very special treatment
for the calculation of its quantities.
The emitted particle, emerging
from the particle collision out of an extremal 
black hole,
can only move for radii for which the square root
in the definition of $Z$,  $Z
= \sqrt{X^2 - m^2 f}$, see Eq.~\eqref{eq:motion_r_penrose}, is well
defined. This means that particle motion can only happen for radii
satisfying $X \geq m \sqrt{f\,\,}$, with 
$X=E - \frac{ e Q}{\left(d-3\right) r^{d-3}}$, see
Eq.~\eqref{eq:motion_r_penrose}.
Thus, $X \geq m \sqrt{f\,\,}$,
translates into
$
e\leq
\frac{\left(d-3\right) r^{d-3}}{Q}
\left(E-m \sqrt{f\,\,}\right)
$.
From the definition of $f$
of an extremal black hole, see Eq.~\eqref{eq:f_factorization},
we have 
$r^{d-3}=\frac{r_+^{d-3}}{(1-
\sqrt{\frac{f}{g}})^{d-3}}$,
so that 
$
e\leq
\frac{\left(d-3\right) r_+^{d-3}}{Q}
\frac{E-m \sqrt{f(r)}}
{
(1-\sqrt{\frac{f(r)}{g(r)}})^{d-3}
}
$, i.e.,
$
e\leq
\frac{\left(d-3\right) r_+^{d-3}E}{Q}
\frac{1-\frac{m}{E} \sqrt{f(r)}}
{
(1-\sqrt{\frac{f(r)}{g(r)}})^{d-3}
}
$.
We identified critical charge $e_c$
as 
$e_c=\frac{\left(d-3\right) r_+^{d-3}E}{Q}$, see
Eq.\eqref{eq:critical_charge_definition}.
So, one has
$
e\leq
e_c
\frac{1-\frac{m}{E} \sqrt{f(r)}}
{
(1-\sqrt{\frac{f(r)}{g(r)}})^{d-3}
}
$. Now for each particle $i$, the function on
the righr hand side of this
inequality acts as a
potential barrier, so we define the potential
${e_i}_0$ for particle $i$ as
$
{e_i}_0(r)\equiv
{e_i}_c
\frac{1-\frac{m_i}{E_i} \sqrt{f(r)}}
{
(1-\sqrt{\frac{f(r)}{g(r)}})^{d-3}
}$,
with ${e_i}_0(r_+)=
{e_i}_c$. For particle 3, $i=3$, we have
$
{e_3}_0(r)\equiv
{e_3}_c
\frac{1-\frac{m_3}{E_3} \sqrt{f(r)}}
{
(1-\sqrt{\frac{f(r)}{g(r)}})^{d-3}
}$ as the potential that particle 3 feels.
Clearly ${e_3}_0(r)$ establishes where 
motion is allowed,
and since $e_3\leq {e_3}_0(r)$ the
allowed motion
depends on the particle's charge.
Since, we are assuming $Q>0$,
we assume that $e_3 > 0$, so that the
criticality condition $X_3(r_+) = 0$ can be satisfied,
bearing in mind that these assumptions are
without loss of generality.
The collisions of interest
occur near the horizon where $f(r)$ is small,
so ${e_3}_0(r)$ can be expanded in $\sqrt{f(r)}$
as
\begin{equation}
{e_3}_0(r)\equiv
{e_3}_c
\left[
1+\left(
\frac{d-3}{\sqrt{g(r)}}
-\frac{m_3}{E_3}
\right)
\,\sqrt{f(r)}\right],
\label{eq:effective_potential_expansion}
\end{equation}
plus $\mathcal{O}\left(f\right)$ terms
and with ${e_3}_0(r_+)=
{e_3}_c$.
This ${e_3}_0(r)$ is the potential that particle 3 feels
at this order of approximation,
and the condition for particle $3$ motion is then
that the electric charge $e_3$
obeys
\begin{equation}
e_3\leq {e_3}_0(r)\,
\label{eq:effective_potential1}
\end{equation}
with ${e_3}_0(r)$
given in Eq.~\eqref{eq:effective_potential_expansion}.
Note that for  particle $3$ resulting from a collision in the
vicinity of the black hole horizon to be able to escape outwards,
one has to have $m_3
< \frac{d-3}{\sqrt{g(r)}}E_3$, such that the term proportional
to $\sqrt{f\,\,}$
in Eq.~\eqref{eq:effective_potential_expansion}
is positive at least close to the horizon.
Otherwise, 
${e_3}_0(r)$ decreases with increasing $r$, and there would be
a radius where the condition
Eq.~\eqref{eq:effective_potential1} would be violated
and motion would not be possible.
Note that  $m_3
< \frac{d-3}{\sqrt{g(r)}}E_3$ can be written as
$E_3>\frac{\sqrt{g(r)}}{d-3}m_3$
so that one can define a lower energy
${E_3}_l\equiv\frac{\sqrt{g(r)}}{d-3}m_3$, such
that 
$E_3>{E_3}_l$.

There are four cases of interest in which collisions can happen in
the vicinity of the horizon.
All of them have $m_3\frac{\sqrt{g(r_+)}}{d-3}\leq
E_3$. We now enumerate the four cases
of this collisional Penrose process.
{\bf 1.} $m_3\frac{\sqrt{g(r_+)}}{d-3}\leq
E_3$, $e_3 \leq {e_3}_c$, and
$\varepsilon_3 = + 1$. The collision happens in the allowed
region, just near the horizon, where $e_3 \leq
{e_3}_c\leq {e_3}_0(r)$.
After the collision, particle 3 moves to larger radii
 since $\varepsilon_3 = + 1$.
Particle 3 is near critical or critical, since
in this case the collision can happen at
the horizon itself. 
{\bf 2.}  $m_3\frac{\sqrt{g(r_+)}}{d-3}\leq
E_3$, ${e_3}_c < e_3<{e_3}_0(r)$,
and $\varepsilon_3 = + 1$. The collision happens in the allowed
region, with the condition $e_3 < {e_3}_0(r)$ meaning that it happens
at some radius just outside the horizon radius $r_+$,
i.e., particle 3 is near-critical. Since $\varepsilon_3 = + 1$,
particle 3 moves outwardly immediately after the collision.
{\bf 3.}
$m_3\frac{\sqrt{g(r_+)}}{d-3}\leq E_3$, $e_3\leq{e_3}_c \leq
{e_3}_0(r)$, and $\varepsilon_3 = - 1$.
The collision happens in the allowed region, just near the horizon,
where $e_3\leq e_{3c} \leq e_{30} (r)$.
Particle 3 is near critical or critical,
since in this case the collision can happen at the horizon itself.
Since $\varepsilon_3 = - 1$, the
particle moves first toward the horizon, reaches a turning point
before it, and moves back in the outward direction. This turning point
is found from Eq.~\eqref{eq:effective_potential1}.  After knowing the
electric charge $e_3$ of the emitted particle 3, one finds from
Eq.~\eqref{eq:effective_potential1} the radius $r$ at which $e_3=
{e_3}_0(r)$, with ${e_3}_0(r)$ given in
Eq.~\eqref{eq:effective_potential_expansion}, with this $r$ being the
turning point.
{\bf 4.} $m_3\frac{\sqrt{g(r_+)}}{d-3}\leq E_3$, ${e_3}_c <
e_3<{e_3}_0(r)$, and $\varepsilon_3 = - 1$. The collision happens in
the allowed region, with the condition $e_3 < {e_3}_0(r)$ meaning that
it happens at some radius $r$ just outside the horizon radius $r_+$,
i.e., particle 3 is near-critical. Since $\varepsilon_3 = - 1$, the
particle moves first toward the horizon, reaches a turning point
before it, and moves back in the outward direction.

To distinguish systematically between cases 1., 2., 3., and 4., 
a parameter
$\delta$ is defined in such a way that
\begin{equation}
e_3 = {e_3}_c \left(1 + \delta\right),
\label{eq:delta}
\end{equation}
where $\delta < 0$ in cases 1. and 4., and $\delta \geq 0$ in
cases
2. and 3. 
For  cases 1., 2., 3., and 4.
the collisions occur at $r$ near the horizon $r_+$, and,
therefore, $X_3$ and $Z_3$ can be expanded for small values of
$r-r_+$ which can then be substituted by $\sqrt{f(r)}$.
Thus, $\delta$ is very small,
and comparing Eq.~\eqref{eq:nearcriticalparticle}
with Eq.~\eqref{eq:delta} we see that particle 3, the emitted
particle, is
indeed a near critical particle.
For near critical particles, $\delta$ should be controlled
and one way to do it is to expand $\delta$ in a series
in $\sqrt{f\,\,}$ which is near zero for $r$ near $r_+$, 
\begin{equation}
   \delta = \frac{\Delta e_3}{e_3}\, \sqrt{f\,\,}  ,
\label{eq:delta_series}
\end{equation}
plus $\mathcal{O}\left(f\right)$ and 
where $\frac{\Delta e_3}{e_3}$ is some constant value,
not infinitesimal,
see \cite{zaslavskii2} for $d
= 4$ and ${k}=0$.
The quantity $\frac{\Delta e_3}{e_3}$
is to be determined from
the conservation equations
as a function of geometrical quantities and
particle quantities.
There is also an upper bound for $\frac{\Delta e_3}{e_3}$,
imposed by the condition $e_3<{e_3}_0(r)$, which
from Eqs.~\eqref{eq:effective_potential_expansion}
and \eqref{eq:delta}
reads $\frac{\Delta e_3}{e_3}
<\frac{d-3}{\sqrt{g(r)}}-\frac{m_3}{E_3}$.
Now, since particle
3 is near critical,
$X_3$ and $Z_3$ can be expanded in $\sqrt{f(r)}$ as
\begin{equation}
X_3 = E_3
\left(\frac{d-3}{\sqrt{g(r)}}
 - \frac{\Delta e_3}{e_3}\right)\,\sqrt{f(r)}\,,\quad\quad
 Z _3= E_3
\sqrt{
\left(
\frac{d-3}{\sqrt{g(r)}}-\frac{\Delta e_3}{e_3}\right)^2
- \frac{m_3^2}{E_3^2}
}\,\sqrt{f(r)}
\,,
\label{eq:XZ3_near_critical}
\end{equation}
plus $\mathcal{O}\left(f\right)$ terms.
For $\delta<0$ one has $\frac{\Delta e_3}{e_3}<0$ and all is fine
in Eq.~\eqref{eq:XZ3_near_critical}.
For $\delta>0$ one gets
from Eq.~\eqref{eq:XZ3_near_critical} the bound
$\frac{\Delta e_3}{e_3}< \frac{d-3}{\sqrt{g(r)}}$
which is weaker than the one
just found above.

From the momentum conservation relation,
Eq.~\eqref{eq:momentum_conservation},
i.e., 
$
\varepsilon_1 Z_1 + \varepsilon_2 Z_2 = \varepsilon_3 Z_3 +
   \varepsilon_4 Z_4$,
together with
electric charge and energy conservation,
Eqs.~\eqref{eq:charge_conservation} and
\eqref{eq:energy_conservation}, respectively, 
   assuming that particle 1 is
critical and particles 2 and 4 are usual,
see Eq.~\eqref{eq:124extremal},
and $\varepsilon_1 =
\varepsilon_2 = \varepsilon_4 = -1$, one finds
upon using Eq.~\eqref{eq:XZ3_near_critical}
that
\begin{equation}
 \frac{ d-3}{\sqrt{g(r_+)}}
\left[E_1 - \sqrt{E_1^2 - m_1^2\,
  \frac{g(r_+)}{\left( d-3\right)^2}}\right]
+ E_3 \left( \frac{\Delta e_3}{e_3} -
 \frac{ d-3}{\sqrt{g(r_+)}}\right)
= \varepsilon_3 E_3\sqrt{\left(
  \frac{ d-3}{\sqrt{g(r_+)}}-
  \frac{\Delta e_3}{e_3}\right)^2 -
  \frac{m_3^2}{E_3^2}}\,,
\label{eq:conditions1}
\end{equation}
valid in order $\sqrt f$. From this equation, the expression for the
energy at the center of mass obtained in Eq.~\eqref{eq:ECM5} can be
recovered, considering that all the particles involved in this process
have the same mass $m$,
see the Appendix~\ref{sec:app_app} for the 
derivation.
One can solve Eq.~\eqref{eq:conditions1}
for $\frac{\Delta e_3}{e_3}$. Before that, the first term
that appears in Eq.~\eqref{eq:conditions1}
is an important quantity
as we are about to find, and so we define
\begin{equation}
\Delta E_1 \equiv
  \frac{d-3}{\sqrt{g(r_+)}}
\left[E_1 - \sqrt{E_1^2 - m_1^2\,
  \frac{g(r_+)}{\left( d-3\right)^2}}\,\,\right].
\label{eq:DeltaE1}
\end{equation}
Then, after
 solving Eq.~\eqref{eq:conditions1}, $\Delta e_3$
 is the expression
$\Delta e_3 =\frac1{E_3}\frac{d-3}{\sqrt{g(r_+)}}
  (
E_3  -  \frac{m_3^2 +
(\Delta E_1)^2}
  {2  \Delta E_1 }
\frac{\sqrt{g(r_+)}}{d-3})\,e_3
$. Clearly,
  the quantity $\frac{m_3^2 +
(\Delta E_1)^2}
  {2  \Delta E_1 }\frac{\sqrt{g(r_+)}}{d-3}$ can be defined as an energy,
  and we define $E_{3b}$, with $b$ for bound,
  as $E_{3b}\equiv\frac{m_3^2 +
(\Delta E_1)^2}
  {2  \Delta E_1 }\frac{\sqrt{g(r_+)}}{d-3}$. Upon using $\Delta E_1$
  of Eq.~\eqref{eq:DeltaE1}, the expression for 
$E_{3b}$ can be put in the form
\begin{equation}
E_{3b}
= \frac{1}{2} \frac{m_3^2+m_1^2}{m_1^2} \, E_1
+ \frac{1}{2} \frac{m_3^2-m_1^2}{m_1^2} \, \sqrt{E_1^2 -
m_1^2 \frac{g(r_+)}{(d-3)^2}}.
    \label{eq:g_of_nu}
\end{equation}
Note that any information about particle 2 has disappeared from
the formulas above, but in fact it has been deposited hiddenly
in the quantities pertaining to particle 3.
Now, from Eqs.~\eqref{eq:conditions1}-\eqref{eq:g_of_nu}
we have that
\begin{equation}
\Delta e_3 =\frac{d-3}{\sqrt{g(r_+)}}
  \left(
1  - \frac{E_{3b}}{E_3}\right)\,e_3\,.
\label{eq:conditions2b}
\end{equation}
From Eq.~\eqref{eq:conditions2b},
we deduce that if 
$E_3 \leq E_{3b}$, then
$\frac{\Delta e_3}{e_3}\leq0$ and from
Eq.~\eqref{eq:delta_series}
one has $\delta\leq0$, and then 
from
Eq.~\eqref{eq:delta}
one has  $e_3 \leq {e_3}_c$.
Then, for this case one can write $e_3 \leq
{e_3}_c\leq{e_3}_0(r)$.
If
$E_3 > E_{3b}$,
then
$\frac{\Delta e_3}{e_3}>0$
 and from
Eq.~\eqref{eq:delta_series}
one has $\delta>0$, and then 
from
Eq.~\eqref{eq:delta}
one has  $e_3 > {e_3}_c$.
Since
$e_3<{e_3}_0(r)$ for sure, one is within the case
${e_3}_c < e_3<{e_3}_0(r)$.
A particular delicate case is when $m_1$ is small, even zero, 
$m_1=0$, as can be seen from Eq.~\eqref{eq:g_of_nu}, since
then  $E_{3b}$ is very large, even infinite,
and one has to decide whether $E_3$ is equal or lower than
$E_{3b}$ or $E_3$ can be higher
than $E_{3b}$, so that 
$\frac{\Delta e_3}{e_3}\leq0$ and there is no turning point
or $\frac{\Delta e_3}{e_3}>0$ and there is a turning point,
respectively.
We now show that
to have a turning point one must have $m_1>0$, i.e., $m_1$
cannot be a massless particle.
For that, we 
expand Eq.~\eqref{eq:conditions2b} for small $\frac{m_1}{E_1}$
to find that for
$\frac{\Delta e_3}{e_3}$
to be positive then one has
$\frac{m_3^2}{E_3}<\frac{m_1^2}{E_1}$.
To have a turning point one has to have
$Z_3\neq X_3$ which from Eq.~\eqref{eq:XZ3_near_critical}
means that $\frac{m_3}{E_3}$ is not zero,
more precisely $\frac{m_3}{E_3}>0$,
therefore also $\frac{m_3}{\sqrt{E_3}}>0$.
Thus, we conclude that
one has $m_1>\sqrt{\frac{E_1}{E_3}}\, m_3$
for having a turning point.
Since the right hand side is never zero, we have 
$m_1>0$ mandatorily, in order to have a turning
point in the case $m_1$ small. So for this case
$m_1=0$ is excluded, i.e., $m_1$ can be very small
but not zero.
In the case that 
$m_1\leq\sqrt{\frac{E_1}{E_3}}\, m_3$ then
$\frac{\Delta e_3}{e_3}\leq0$ and there is no turning point,
in particular $m_1=0$ has no turning point.
There is still a further important equation
that we have to explicitly give.
Indeed, Eq.~\eqref{eq:conditions1}
together with Eq.~\eqref{eq:conditions2b}
gives
\begin{equation}
  \varepsilon_3 \lvert(\Delta E_1)^2 - m_3^2\rvert=
(\Delta E_1)^2 - m_3^2.
\label{eq:conditions3}
\end{equation}
If $\Delta E_1>m_3$ we deduce from Eq.~\eqref{eq:conditions3}
that  $\varepsilon_3=+1$, particle 3
goes out from the point of collision.
If $\Delta E_1<m_3$ we deduce from Eq.~\eqref{eq:conditions3}
that $\varepsilon_3=-1$, particle 3
goes in from the point of collision,
in particular when $m_1$ is sufficiently small
and so $\Delta E_1$ is small,
which comprises the case $m_1=0$, particle 3 goes in.
If $\Delta E_1=m_3$, Eq.~\eqref{eq:conditions3} is an identity.

We have still to analyze the energy of particle 4, since to
have energy extraction and
in order that the BSW collision gives rise to a Penrose
process, this energy has to be negative.
Since the electric charge $Q$
of the black hole is the same for all four particles,
the point $r$ of collision is the same for all four particles,
and there is conservation of electric charge, i.e., $e_1+e_2-e_3-e_4=0$,
see Eq.~\eqref{eq:charge_conservation},
we have that the conservation equation
$X_1 + X_2 = X_3 + X_4$, 
see Eq.~\eqref{eq:energy_conservation},
implies that
at the point of collision, $E_1 + E_2 = E_3 + E_4$,
i.e., $E_4=  E_1+E_2 -E_3$.
For energy extraction, one necessarily has
$E_3>E_1+E_2$.
Thus, when
there exists extraction one finds
$E_4<0$
necessarily, and so $e_4<0$.
Therefore, when
there is energy extraction
there is an electric ergosphere 
for particle 4, 
characterized by 
$r_+\leq r<
r_{\mathrm{ergo}\,4}$ with
$r_{\mathrm{ergo}\,4}^{d-3} =\frac{\lvert e_4 \rvert Q}{\left(d-3\right)
\lvert E_4 \rvert }$, see Eq.~\eqref{eq:ergosphere},
see also Appendix~\ref{sec:app_app}.
There is thus 
a Penrose process or, when $E_3\to\infty$, one
is in the presence of a super-Penrose process.

From the analysis made, namely,
from the conditions that follow from 
Eqs.~\eqref{eq:conditions2b}
and \eqref{eq:conditions3}
and the discussions after 
them, the four cases
considered above can be spelled
out in detail,  with the calculated
bounds for the energy and mass of particle 3,
the emitted particle.
The four cases mentioned
of the collisional Penrose process we are
studying can be now enumerated
in full detail 
as follows.
\begin{enumerate}
  \item $\mathrm{OUT}^-$:
   $0\leq m_3 < \Delta E_1$ and
   $m_3\frac{\sqrt{g(r_+)}}{ d-3} \leq E_3 \leq E_{3b}$.
   Thus, $e_3 \leq {e_3}_c\leq{e_3}_0(r)$ and $\varepsilon_3=+1$,
   the particle goes directly out after the collision.
   This case, yields
   no energy extraction since $E_3$ is less than $E_1$. This
   happens because, as $m_3 < \Delta E_1$ one has $E_{3b}< \Delta E_1
   \frac{\sqrt{g(r_+)}}{d-3} $, which is lower than $E_1$, from
   Eq.~\eqref{eq:DeltaE1}. 
\item $\mathrm{OUT}^+$:
  $0\leq m_3 < \Delta
   E_1$ and $E_{3b}
   \leq E_3<\infty$.
 Thus, ${e_3}_c < e_3<{e_3}_0(r)$ and $\varepsilon_3=+1$,
   the particle goes directly out after the collision.
This case, can yield a Penrose process with
  energy extraction and it is not restricted, but there is an upper
  bound in $m_3$, i.e., not any mass can be emitted. Moreover,
  clearly, when $m_1$ is very small  then $E_{3b}$
   can be arbitrarily large but not
   infinite, and the energy extracted in particle 3, $E_3$,
   can be arbitrarily large but not infinite as well,
   characterizing thus a super-Penrose process. When there
   is energy extraction one has $E_4<0$ and there is an
   electric ergosphere.
\item $\mathrm{IN}^-$:
  $\Delta E_1<m_3<\infty$ and 
  $m_3\frac{\sqrt{g(r_+)}}{ d-3} \leq E_3
   \leq E_{3b}$.
 Thus, $e_3 \leq {e_3}_c\leq{e_3}_0(r)$ and $\varepsilon_3=-1$,
 the particle goes in immediately  after the collision and then
   continues the motion entering down the black hole.
We have seen that for the case 
$m_1\leq\sqrt{\frac{E_1}{E_3}}\, m_3$ there is no turning point
for the ingoing particle.
So, small masses $m_1$ which yield high $E_{3b}$
have no turning points.
There is no extraction of energy at all.
In particular, for $m_1=0$,
which falls within this case, one obtains $E_{3b}=\infty$,
and therefore $E_3=\infty$, but all this energy
goes down the black hole.
\item $\mathrm{IN}^+$:
  $\Delta E_1<m_3<\infty$ and
  $ E_{3b}
   \leq E_3<\infty$.
 Thus, ${e_3}_c < e_3<{e_3}_0(r)$ and $\varepsilon_3=-1$,
   the particle goes in immediately after the collision and then
   reverses the motion at some radius nearer the horizon
   to move outward from then on.
This case, can yield a Penrose process with
  energy extraction, it is not restricted and there is a lower bound
  in $m_3$. Moreover,
  clearly, when $m_1$ is small then $E_{3b}$
   can be arbitrarily large
   but not infinite, and the energy extracted in particle 3, $E_3$,
   can be arbitrarily large but no infinite as well,
   characterizing thus a super-Penrose process.
   There is also the possibility that
   $m_3$ be very large in this case,
   allowing thus for the emission of superheavy particles.
   When there
   is energy extraction one has $E_4<0$ and there is an
   electric ergosphere.
\end{enumerate}

To have a hand on these results and what one gets out of them, let us
suppose that the quantities given, the initial inputs, are $m_1$,
$E_1$, $e_1$, $\varepsilon_1$, $m_2$, $E_2$, $e_2$, $\varepsilon_2$,
and the point $r$ of the collision of particles 1 and 2. This point of
collision, to be of interest, is near the horizon radius $r_+$ and can,
in principle, 
be directly found from kinematic expressions for particles 1 and
2.  It is also assumed that particle 4 travels inward so that
$\varepsilon_4$ is known, plus that particle 4 is a usual particle.
Then, $m_3$, $E_3$, $e_3$, and $\varepsilon_3$ can have different
values, depending on the collision internal process itself. One has
that $\varepsilon_3$ can be either $+1$ or $-1$.
For each
collision,
one could consider 
$m_3$, $E_3$, and $e_3$ as free parameters, 
but this cannot be, since
for the given inputs above,
$m_3$, $E_3$, and $e_3$ are entangled quantities, as can be
seen from Eqs.~\eqref{eq:delta},
\eqref{eq:delta_series}, and \eqref{eq:conditions2b}.
Let us then consider $m_3$ and $E_3$ as free parameters.
Thus, $m_3$ emitted can have some range of values
and $E_3$ emitted is also 
within some range, but within these ranges
$m_3$ and $E_3$ can be any, they are going to
depend on the very details of the collision internal process.
Given $m_3$ and $E_3$, the electric charge $e_3$ is then
fixed, with its allowed values being also within some range.
Now, in the
case particle 3 moves always inward then it is hard to measure $m_3$
and $E_3$, but this is anyway irrelevant since it is the case
$\mathrm{IN}^-$ and there is no energy extraction.  In the case
particle 3 moves outward in one stage or the other, which is the
interesting case, we need a detector to measure $m_3$ and $E_3$, with
$e_3$ being then calculated from Eqs.~\eqref{eq:delta} and
\eqref{eq:delta_series}.  From the measured value of $m_3$ and the
initial inputs, we find $\varepsilon_3$, making use of
Eq.~\eqref{eq:conditions3}, and so we can discern between the cases
$\mathrm{OUT}$, be it $\mathrm{OUT}^-$ or $\mathrm{OUT}^+$, that has
$\varepsilon_3=+1$, and $\mathrm{IN}^+$ that has $\varepsilon_3=-1$,
this latter case yielding directly energy extraction. To distinguish
between $\mathrm{OUT}^-$ and $\mathrm{OUT}^+$ we have to calculate
$E_{3b}$ through Eq.~\eqref{eq:g_of_nu}.  Given $m_1$ and $E_1$, and
measuring $m_3$, one calculates $E_{3b}$.  Then, if the measured $E_3$
is in the range $m_3\frac{\sqrt{g(r_+)}}{ d-3} \leq E_3 \leq E_{3b}$
one is in the case $\mathrm{OUT}^-$, and there is no energy
extraction.  If the measured $E_3$ is in the range $E_{3b} \leq
E_3<\infty$ one is in the case $\mathrm{OUT}^+$, and there is energy
extraction.  Moreover, we can find from the conservation laws,
Eqs.~\eqref{eq:energy_conservation}-\eqref{eq:charge_conservation},
the other physical quantities of
particle 4, i.e., $m_4$, $E_4$, and $e_4$, given the inputs and the
measured values of $m_3$ and $E_3$.
Note that since there are many quantities, namely,
 $m_1$,
$E_1$, $e_1$, $\varepsilon_1$, $m_2$, $E_2$, $e_2$, $\varepsilon_2$,
 $m_3$,
$E_3$, $e_3$, $\varepsilon_3$, $m_4$, $E_4$, $e_4$, $\varepsilon_4$,
and the point of collision $r$,
what is initial input, what is measured, and what is deduced
is a matter of choice. Above, we have given
what we think is an interesting example
of a collisional Penrose process in which
the collisions are of BSW type. But a great number
of other examples might be given.
For instance, if in the example above, the detector is also able to
measure the electric charge $e_3$, then the point of collision
does not need to be an initial input, it can be found a posteriori
through  Eqs.~\eqref{eq:delta_series} and
\eqref{eq:conditions2b}.

\subsection{The dependence on the cosmological constant through
${k}$ and on
the dimension $d$ and further comments}

\subsubsection{The dependence on the cosmological constant through
${k}$ and on
the dimension $d$}

The bounds for the energy extracted, i.e., the energy of particle 3,
depend on the factor $\frac{\sqrt{g(r_+)}}{d-3}$,
see Eq.~\eqref{eq:g_of_nu}.  Using
the factorization of $f\left(r\right)$ obtained in
Eq.~\eqref{eq:f_factorization}, it follows that
$
\frac{g(r_+)}{(d-3)^2}
=\frac12
 \frac{r_+^2}{\left(d-3\right)^2} \frac{d^2f}{dr^2}(r_+)
$.
The second derivative of $f\left(r\right)$ can be computed from
Eq.~\eqref{eq:line_element}, leading to
$\frac12\frac{d^2f}{dr^2}(r_+)=
 -\left(d-3\right)\left(d-2\right) \frac{\mu \ \, M}{r_+^{d-1}} + 
 \left(d-3\right)\left(2d-5\right) \frac{\chi
 Q^2}{r_+^{2\left(d-2\right)}} -  \frac{{k}}{l^2}$.
Using this result in the previous
equation, it follows that
$
\frac{g(r_+)}{(d-3)^2}
=
- \frac{d-2}{d-3} \frac{\mu
 \, M}{r_+^{d-3}}
 + \frac{2d-5}{d-3}
 \frac{\chi Q^2}{r_+^{2\left(d-3\right)}}
-{k} \,
 \frac{1}{\left(d-3\right)^2}\,\frac{r_+^2}{l^2}
$.
Imposing the condition for an extremal black hole, Eq.~\eqref{r+},
the following result can finally be obtained
\begin{equation}
\frac{g(r_+)}{(d-3)^2} = 1 - {k} \frac{\left(d-2\right)
 \left(d-1\right)}{\left(d-3\right)^2}\,
 \frac{r_+^2}{l^2} .
\label{eq:factor_final_result}
\end{equation}
From Eq.~\eqref{eq:factor_final_result} we can discuss
the dependence of the results on the spacetime dimension $d$
and also on the cosmological constant $\Lambda$, i.e.,
the cosmological length $l$.

Firstly, we analyze the dependence on $d$. 
For ${k}=-1$, one has that $\frac{g(r_+)}{(d-3)^2}$
increases in the presence of the cosmological constant
and depends on $d$.
For ${k} = 0$, $\frac{g(r_+)}{(d-3)^2} = 1$,
the bounds do not depend on the number of
dimensions $d$, and the
bounds are the same obtained for an
asymptotically flat Reissner-Nordstr\"om black hole spacetime in
$d=4$ \cite{zaslavskii2}. For ${k}=1$, the factor
$\frac{g(r_+)}{(d-3)^2}$
decreases due
to the presence of the cosmological constant,
noting also that $\frac{\left(d-2\right)
 \left(d-1\right)}{\left(d-3\right)^2}\,
 \frac{r_+^2}{l^2}<1$, since the $r_+$ we are considering
 is the black hole horizon, not the cosmological one, and the
bound depends on $d$.
Of course, one could redefine the cosmological
constant, i.e., the cosmological length $l$,
to include the $d$-dependent factors, but
surely in such a case other quantities, not necessarily
related to this problem,
that depend on the pure cosmological constant 
would then depend on the dimension $d$
through the inverse of the  factor 
 $\frac{\left(d-2\right)
 \left(d-1\right)}{\left(d-3\right)^2}$.

Secondly, we analyze the dependence on the cosmological
constant $\Lambda$, i.e.,
the cosmological length $l$.
The bound $E_{3b}$
for the energy of the emitted particle can be interpreted as
function of $\frac{\sqrt{g(r_+)}}{d-3}$ given by
$
E_{3b}
= \frac{1}{2} \frac{m_3^2+m_1^2}{m_1^2} \, E_1
+ \frac{1}{2} \frac{m_3^2-m_1^2}{m_1^2} \, \sqrt{E_1^2 -
m_1^2 \frac{g(r_+)}{(d-3)^2}}
$, see Eq.~\eqref{eq:g_of_nu}.
Therefore, for the cases
for which one has a collisional Penrose process
and one can have energy extraction, i.e.,
$\mathrm{OUT}^+$ and $\mathrm{IN}^+$,
one has to consider different possibilities.
First, for $k=0$ the lower bound $E_{3b}$
is independent of the cosmological
constant, directly from Eq.~\eqref{eq:factor_final_result},
together with Eq.~\eqref{eq:g_of_nu}, and of $d$.
Thus, we can compare the cases of negative
cosmological constant, i.e., $k=-1$,
with the cases of 
positive
cosmological constant, i.e., $k=1$, for
fixed $\frac{r_+}{l}$, i.e., fixed
horizon radius relative to the
cosmological radius, which means
compare $\frac{r_+}{l}$ in the various situations.
If $m_3 < m_1$, the lower
bound $E_{3b}$
for the energy of the emitted particle
is greater for negative cosmological constant, i.e.,
$k=-1$, than 
 for positive cosmological constant, i.e.,
$k=+1$.
If $m_3 = m_1$, the lower
bound $E_{3b}$
for the energy of the emitted particle
is equal for negative cosmological constant, i.e.,
$k=-1$, and
 for positive cosmological constant, i.e.,
$k=+1$, which means that $E_{3b}$ does not depend on the
cosmological constant.
If $m_3 > m_1$, the lower
bound $E_{3b}$
for the energy of the emitted particle
is smaller for negative cosmological constant, i.e.,
$k=-1$, than 
 for positive cosmological constant, i.e.,
$k=+1$.
An interpretation can be tried. Define $\Delta m\equiv m_3-m_1$.
One can associate a net  lengthscale
$\lambda$ by $\lambda=\frac1{\Delta m}$, with
$\lambda$
pointing inward, i.e., to decreasing radii,
if  $\Delta m<0$, and 
$\lambda$
pointing outward, i.e., to increasing radii,
if  $\Delta m>0$.
In this convention one also has that the AdS
lengthscale $l$ points inward, 
and the dS
lengthscale $l$ points outward.
Thus, since for $\Delta m<0$, $\lambda$ net points inward,
$l$ AdS points inward and $l$ dS points outward,
one needs higher energy, higher $E_{3b}$, in 
AdS in relation to dS to have the particle going out.
Also, since for $\Delta m>0$, $\lambda$ net points outward,
$l$ AdS points inward and $l$ dS points outward,
a particle that can go out, goes  out with lower energy,
lower $E_{3b}$, in 
AdS in relation to dS.
For $\Delta m=0$ one has that $\lambda$ net is infinite
and does not point in any direction,
and so there is no dependence
on the cosmological constant.
This is the best we can offer as an interpretation.

In brief, different conclusions can be obtained depending on the
parameter ${k}$, i.e., on the sign of the cosmological constant. For
${k}=0$, i.e., for an asymptotically flat Reissner-Nordstr\"om black
hole spacetime, the bounds for the energy extracted
in the collisional Penrose processes considered do not depend on
the number of dimensions. For ${k} = \pm 1$, these bounds depend on
the number of dimensions, on the sign of ${k}$, and
on the masses $m_1$ and
$m_3$. These bounds can be larger or lower than for ${k}=0$ depending
on the sign of ${k}$ and on whether $m_1$ is lower or greater than
$m_3$.

We have considered the possible scenarios for the collisional Penrose
process from a BSW collision, of colliding electrically charged
particles in a Reissner-Nordstr\"om background with cosmological
constant and in $d$ dimensions.  One could think of considering of
doing a classification of the decayment Penrose process, instead of
the collisional Penrose process, to find the possible scenarios in the
generic background considered here. This decayment Penrose process of
electrically charged particle classification has been done in a
Reissner-Nordstr\"om background with zero cosmological constant in
$d=4$ dimensions \cite{zaslavskii2024}.  A detailed comparison of
these two types of Penrose processes, the collisional and the
decayment, is certainly of great interest.

\subsubsection{Further comments}

In the collision processes that we have studied, an essential role is
played by critical or near-critical particles in the vicinity of the
black hole horizon.  Such particles can be obtained by fine-tuning
their parameters in their initial state, i.e., in the state previous
to the collision.  One might wonder how does such type of particles
come into being in a plausible manner.  We can think of two ways. One
is through an experiment where the physicist prepares the particles in
the necessary fine tuned manner.  The other is to have have a cluster
of a myriad of particles with all sorts of energy in the vicinity of
the black hole such that statistically some of them are indeed near
critical.

Another question that can be raised is how the collisions in the
vicinity of an extremal black hole horizon can give rise to measurable
physical observables.  In our study, we have been interested in the
understanding of the energetics of the possible processes.  We have
discussed not only the local center of mass energies $E_{\rm cm}$ in
the collisions, which is indeed the original BSW effect, but have gone
beyond it and discussed the Killing energies $E$ that can be achieved
in the collision.
We have found that there are situations in which the energy $E$ 
is larger than the initial energy characterizing a collisional
Penrose process. In addition,
we have found that there are situations that not
only the center of mass energies $E_{\rm cm}$ are arbitrarily high but
also the Killing energies are arbitrarily large. We have thus displayed
examples of  super-Penrose processes.  If particles have arbitrarily
large Killing energies it means they can reach far
distances, although
impediments to it can arise. For instance, the particles ejected can
lose their energy through scattering processes or even fall down the
black hole. Thus, independently of whether or not the ejected
particles reach far distances, the issue of the measuring of physical
observables by an asymptotic observer, is worth pursuing, although it
has not been touched by us.  Nevertheless, we can make now some
comments related to the electrically charged case which is the one we
have studied. On could think of direct and indirect detections.
Direct detections are indeed possible here.  Since the outgoing
charged particle can have arbitrarily large energy, it can be detected
directly at infinity by a charge counter.  One thing we can say for
sure in this direct detection case is that
if the outgoing particle is
directly detected then it is automatically a nearly fine tuned
particle, as a usual outgoing particle cannot escape from a black hole
horizon.
On the other hand,
indirect detections can happen from various radiative
processes involving the outgoing particle interacting
electromagnetically with other particles.  One possible indirect
observational signature could be provided by inverse Compton
scattering, whereby the outgoing highly energetically particle loses
energy when it encounters a photon, which photon can then be detected
at infinity in the gamma ray range.  Another possible observational
signature is through bremsstrahlung, where the outgoing highly
energetic particle is deflected by some other electric charged
particle, loses energy and radiates photons. Bremsstrahlung has a
characteristic spectrum, a continuous spectrum peaking at higher and
higher frequencies as the deceleration of the particle increases.  It
would be of interest to find how these phenomena are affected for
processes with or without electric charge, rotation, or cosmological
constant.

A remark is in order.
Electrically charged black holes of microscopic or macroscopic size
are prone to be discharged by the surrounding medium.
On the other hand, if the black holes are isolated, i.e., with no
surrounding medium, quantum effects come into play.
Due to vacuum polarization near the event horizon of the black hole,
electrically charged isolated black holes are susceptible of
discharging themselves sooner or later. Indeed, an electrically
charged black hole favors the absorption of particles with opposite
charge, losing therefore its charge.  One can think of ways of
bypassing this issue. One can take the charge to be a topological
charge so that there are no particles to radiate. Or else, one can
admit that the lightest charged particle of the theory is massive
enough that cannot be created, and thus cannot be absorbed by the
black hole. For instance, a magnetically charged black hole can lose
its charge only by creating magnetic monopoles, which in principle are
sufficiently massive such that the probability of creation is highly
suppressed and stabilizes the discharging of the black hole. An
alternative to this scheme would be to posit a central charge that
arises in the algebra of supergravity theories.
Interesting to note that for isolated black holes with sufficient high
electric charge, the emission of charged massive particles is
exponentially suppressed due to the Schwinger effect. Such black holes
have to be large and massive, which implies a low electric field at
the event horizon, quenching pair creation of massive particles.  If
in the electromagnetic theory in use, there are no massless charged
particles, as is our case, then the black hole cannot create and emit
the lightest possible massive particle.  These large isolated black
holes emit thus only neutral particles via Hawking radiation and tend
to an extremal state, which are the ones that interest us here.
Thus, the effects we have been studying in this work could arise in
isolated electric charged black holes where suppression effects for
quantum particle creation can occur to stabilize the
electric charge of the black holes, be
they microscopic or macroscopic.
The astrophysical black holes that so far have been observed are
macroscopic objects, which being surrounding by matter, would
discharge quickly, and so for them, rotational effects are the
important ones. We have not dealt with angular momentum effects of a
rotating black hole, nevertheless, the effects we have found for
electrically charged black holes may serve as a guide to black holes
with angular momentum charge.

\section{Conclusions}
\label{sec:concl}

We have analyzed the BSW mechanism
and the corresponding energy extraction in a collisional
Penrose process, 
for a $d$-dimensional extremal
black hole spacetime with horizon radius
$r_+$, electric charge $Q$, and
cosmological constant ${k} \Lambda$, with ${k} = -1, 0, +1$, i.e., the
spacetime can be asymptotically AdS, flat, or dS, respectively.
By identifying that the relevant collision process, namely, the one
that yields divergingly high center of mass energies in the collision
of two ingoing particles, is the collision between a critical and a
usual particle, we set on to study this process in all detail.  In
this, we have put bounds on the mass, energy, and electric charge of
the particle emitted and have classified the cases in which a net
large energy extraction can be obtained from the extremal
Reissner-Nordstr\"om black hole.  In some cases one finds that a
super-Penrose process is possible, where an arbitarily large Killing
energy can be carried by the outgoing particle.  We have also shown
that the created particle that falls down the hole has negative energy
and is surrounded by an electric ergosphere allowing thus the
existence of a Penrose process.  Examples for the particle physical
quantities that are given a priori, what might be measured in the
collision process, and which information is obtained a posteriori can
be made concrete, as was exemplified in one case for particles 1, 2,
3, and 4.
We have shown that the bounds on the energy of the emitted particle do
not depend on the dimension $d$ for zero cosmological constant, i.e.,
for asymptotically flat black hole spacetimes.  On the other hand, the
bounds do depend on $d$ for nonzero cosmological constant, i.e., for
negative cosmological constant or asymptotically AdS spacetimes, and
for positive cosmological constant or asymptotically dS spacetimes.
This dependence can be seen from the expressions for
$\frac{g(r_+)}{(d-3)^2}$ and $E_{3b}$ above, and turns out to be a
weak dependence, from a factor 6 when $d=4$ to a factor 1 when $d$ is
infinite.
We have also shown that the bounds for the energy extracted for each
fixed dimension, are different depending on whether the
cosmological constant is negative, zero, or positive.

\acknowledgments{We acknowledge financial support from Funda\c c\~ao
para a Ci\^encia e Tecnologia - FCT through the
project~No.~UIDB/\break 00099/2020.}

\appendix*

\section{Calculation supporting the results of the main text}
\label{sec:app_app}

\subsection{Getting back the center of mass energy expression}

In Sec.~\ref{sec:E_emitted}, we considered that the four particles
have different masses between themselves and then obtained equations
for the energy of the emitted particle. Now, we show that the
expressions obtained in
Sec.~\ref{sec:E_emitted} lead to the center of mass energy expression in
the particular case that all masses are equal found in Eq.~\eqref{eq:ECM5}
of 
Sec.~\ref{sec:ECM}.

We start by writing the energy at the center of mass frame in terms of
the relevant quantities of particles 3 and 4,
assuming $m\equiv m_1=m_2=m_3=m_3$. This gives
\begin{equation}
\frac{E_{\mathrm{CM}}^2(r)}{2m^2} =
1 + \frac{X_3(r) X_4(r) + \varepsilon_3 Z_3(r) Z_4(r)}{m^2f(r)}\,,
\label{eq:ECM_app_1}
\end{equation}
where we have put $\varepsilon_4=-1$ as was assumed,
and since particle 3 can move in or out, we have left
$\varepsilon_3$ unspecified, it can be $-1$ or $+1$.
Then, from the expressions for $X_3\left(r\right)$,
$X_4\left(r\right)$, $Z_3\left(r\right)$, and $Z_4\left(r\right)$
given in
Eqs.~\eqref{eq:124extremal} and \eqref{eq:XZ3_near_critical},
one finds
\begin{equation}
\frac{E_{\mathrm{CM}}^2(r)}{2m^2} = 1 + \frac{X_4
\left(r_+\right)}{\sqrt{f\left(r\right)} \, m^2} \left[E_3
\left(\frac{d-3}{\sqrt{g\left(r_+\right)}}-\frac{\Delta
e_3}{e_3}\right) + \varepsilon_3 E_3
\sqrt{\left(\frac{d-3}{\sqrt{g\left(r_+\right)}}-\frac{\Delta
e_3}{e_3}\right)^2-\frac{m^2}{E_3^2}}\right]\,.
\label{eq:ECM_app_2}
\end{equation}
Since $X_1\left(r_+\right) = X_3\left(r_+\right) = 0$, we have
from energy conservation, Eq.~\eqref{eq:energy_conservation}, that
$X_4\left(r_+\right) = X_2\left(r_+\right) = E_2 - \frac{ e_2
Q}{\left(d-3\right) r_+^{d-3}}$, and thus
\begin{equation}
X_4\left(r_+\right) = E_2 - \frac{ e_2
Q}{\left(d-3\right) r_+^{d-3}}\,.
\label{eq:X4app}
\end{equation}
From Eq.~\eqref{eq:conditions1} one finds that 
in the case all masses are equal, the following expression
holds
\begin{equation}
\frac{ d-3}{\sqrt{g(r_+)}}
\left[E_1 - \sqrt{E_1^2 - m^2\,
 \frac{g(r_+)}{\left( d-3\right)^2}}\right]
+ E_3 \left( \frac{\Delta e_3}{e_3} -
 \frac{ d-3}{\sqrt{g(r_+)}}\right)
= \varepsilon_3 E_3\sqrt{\left(
 \frac{ d-3}{\sqrt{g(r_+)}}-
 \frac{\Delta e_3}{e_3}\right)^2 -
 \frac{m^2}{E_3^2}}\,.
\label{eq:conditions1app}
\end{equation}
Substituting Eqs.~\eqref{eq:X4app} and \eqref{eq:conditions1app} into
Eq.~\eqref{eq:ECM_app_2} yields
\begin{equation}
\frac{E_{\mathrm{CM}}^2(r)}{2m^2} =
1 + \frac{E_2 - \frac{ e_2 Q}{\left(d-3\right) r_+^{d-3}}}{m^2 \,
g\left(r_+\right)} \left[E_1
-\sqrt{E_1^2-\frac{m^2}{\left(d-3\right)^2} g\left(r_+\right)}\right]
\frac{d-3}{1-\frac{r_+}{r}}\,,
\label{eq:ECM_app_4}
\end{equation}
which is precisely
the expression given in Eq.~\eqref{eq:ECM5}.

\subsection{The ergosphere of particle 4}

In Sec.~\ref{sec:E_emitted} we stated that particle 4 has
negative energy when there is energy extraction
carried out by particle 3. Here we give the details
supporting the statement.
To have energy extraction, it is necessary
that particle 3 has 
energy greater than the initial energy,
the energy just before the collision
of the two initial particles, i.e.,
\begin{equation}
E_3>E_1+E_2.
\label{eq:E3greater}
\end{equation}
At the point of collision the equality 
$E_1 + E_2 = E_3 + E_4$ holds.
This equality comes from 
 Eq.~\eqref{eq:energy_conservation},
$X_1 + X_2 = X_3 + X_4$
together with Eq.~\eqref{eq:motion_r_penrose},
if we use
that the electric charge $Q$
of the black hole is the same for all four particles,
the point $r$ of collision is the same for all four particles,
and there is conservation of electric charge, i.e., $e_1+e_2-e_3-e_4=0$,
see Eq.~\eqref{eq:charge_conservation}.
Thus, from $E_4=  E_1+E_2 -E_3$ and
the necessity of Eq.~\eqref{eq:E3greater}, one finds
\begin{equation}
E_4<0,
\label{eq:E4negative}
\end{equation}
if there is energy extraction in the collision process.
From the forward in time condition $\dot{t} > 0$
for processes outside the horizon applied to particle 4, 
one has $X_4 > 0$, see Eq.~\eqref{eq:motion_t_penrose},
and so one finds
$E_4 -\frac{ e_4 Q}{\left(d-3\right) r^{d-3}}>0$,
where $r$ is the radius of the point of collision
see Eq.~\eqref{eq:motion_r_penrose}.
For energy extraction Eq.~\eqref{eq:E4negative}
holds, so one can write
$-\lvert E_4\rvert -\frac{ e_4 Q}{\left(d-3\right) r^{d-3}}>0$.
Assuming without loss of generality
$Q>0$ as we do, for the latter equation to be true
one necessarily has 
\begin{equation}
e_4<0,
\label{eq:e4negative}
\end{equation}
when there is energy extraction.
Thus, $X_4 > 0$ can be written as 
$-\lvert E_4\rvert +\frac{ \lvert e_4\rvert Q}{\left(d-3\right) r^{d-3}}>0$.
This equation defines a region within which
the inequality is valid, namely 
\begin{equation}
r_+\leq r<
r_{\mathrm{ergo\, 4}}\,,\quad\quad\quad\quad\quad\quad
r_{\mathrm{ergo}\, 4}^{d-3} =\frac{\lvert e_4 \rvert Q}{\left(d-3\right)
\lvert E_4 \rvert },
\label{eq:ergosphere4}
\end{equation}
which is the electric 
ergosphere for particle 4, and is the region
where the collision has to take place
in order that energy extraction can occur.

It is also interesting to note the following.
For the two cases in which energy extraction can occur, i.e.,
$\mathrm{IN}^+$ and $\mathrm{OUT}^+$, see Sec.~\ref{sec:E_emitted},
the electric charge
of particle 3 obeys
the equation
$e_3 > {e_3}_c$, where from Eq.~\eqref{eq:critical_charge_definition}
one has 
${e_3}_c = \frac{r_+^{d-3} \left(d-3\right)}{Q}\,E_3$,
and its energy obeys $E_3
> E_{3b}$. Combining these
two equations one finds
$e_3 > \frac{r_+^{d-3}}{Q} E_{3b}$.
Since we found that
$E_{3b}$ can be as large as one wants, $e_3$ can also assume
arbitrarily large values. Using charge conservation,
Eq.~\eqref{eq:charge_conservation}, one concludes that when
energy extraction occurs, the electric charge of particle 4 becomes
negative, $e_4<0$,
because the energy of the emitted particle grows, as $e_4 =
e_1 + e_2 - e_3$ and $e_3$ becomes dominant.
Then, to have energy extraction $E_4<0$, and so 
$-\lvert E_4\rvert
+\frac{ \lvert e_4\rvert Q}{\left(d-3\right) r^{d-3}}>0$,
and there can be energy extraction if
the collision occurs within the electric ergosphere
defined in
Eq.~\eqref{eq:ergosphere4}.


\end{document}